\documentclass[aip,jcp,preprint]{revtex4-1}%
\usepackage{amsmath,amsfonts,amsthm}
\usepackage{placeins}
\usepackage{dcolumn}
\usepackage{tabularx,multirow}
\usepackage[T1]{fontenc}
\usepackage{graphicx}
\usepackage[english]{babel}
\usepackage{amsmath}
\usepackage{amsfonts}
\usepackage{amssymb}
\usepackage{xcolor}
\usepackage{epstopdf}
\providecommand{\U}[1]{\protect\rule{.1in}{.1in}}
\graphicspath{{./Figures/}{D:/konstantin/QI_Suzuki_extended/Figures/}{C:/Users/Jiri/Dropbox/Papers/Chemistry_papers/2014/Suzuki_QI/QI_Suzuki_extended/Figures/}}
\makeatletter
\providecommand{\tabularnewline}{\\}

\@ifundefined{textcolor}{}
{
\definecolor{BLACK}{gray}{0}
\definecolor{WHITE}{gray}{1}
\definecolor{RED}{rgb}{1,0,0}
\definecolor{GREEN}{rgb}{0,1,0}
\definecolor{BLUE}{rgb}{0,0,1}
\definecolor{CYAN}{cmyk}{1,0,0,0}
\definecolor{MAGENTA}{cmyk}{0,1,0,0}
\definecolor{YELLOW}{cmyk}{0,0,1,0}
}
\makeatother
\begin{document}
\title{Accelerating quantum instanton calculations of the kinetic isotope effects}
\author{Konstantin Karandashev}
\author{Ji\v{r}\'{\i} Van\'{\i}\v{c}ek}
\email{jiri.vanicek@epfl.ch}
\affiliation{Laboratory of Theoretical Physical Chemistry, Institut des Sciences et
Ing\'{e}nierie Chimiques, Ecole Polytechnique F\'{e}d\'{e}rale de Lausanne
(EPFL), CH-1015, Lausanne, Switzerland}
\date{19 November 2015}

\begin{abstract}
Path integral implementation of the quantum instanton approximation currently
belongs among the most accurate methods for computing quantum rate constants
and kinetic isotope effects, but its use has been limited due to the rather
high computational cost. Here we demonstrate that the efficiency of quantum
instanton calculations of the kinetic isotope effects can be increased by
orders of magnitude by combining two approaches: The convergence to the
quantum limit is accelerated by employing high-order path integral
factorizations of the Boltzmann operator, while the statistical convergence is
improved by implementing virial estimators for relevant quantities. After
deriving several new virial estimators for the high-order factorization and
evaluating the resulting increase in efficiency, using $\mathrm{\cdot
H_{\alpha}+H_{\beta}H_{\gamma}\rightarrow H_{\alpha}H_{\beta}+\cdot H_{\gamma
}}$ reaction as an example, we apply the proposed method to obtain several
kinetic isotope effects on $\mathrm{CH_{4}+\cdot H\rightleftharpoons\cdot
CH_{3}+H_{2}}$ forward and backward reactions.

\end{abstract}
\maketitle

\section{Introduction}

Accurate evaluation of the rate constant, i.e., the prefactor of the rate law
of elementary chemical reactions, remains one of the central goals of chemical
kinetics because this constant reflects the mechanism of the reaction as well
as other properties of the potential energy surface on which the reaction
occurs. Another quantity that is frequently used for studying reaction
mechanisms, and, in particular, detecting nuclear quantum effects on reaction
rates, is the kinetic isotope effect (KIE). The KIE is defined as the ratio of
rate constants for two isotopologs, i.e., molecules that only differ in
isotope composition. These effects, which include, e.g., tunneling,
corner-cutting, and zero-point energy effect, tend to play an important role
in hydrogen transfer reactions with a high activation barrier. Although they
are most important at low temperatures, nuclear quantum effects sometimes
manifest themselves even at physiological temperatures, a fact uncovered by
studying KIE's on some enzymatic reactions.\cite{Sen_Kohen:2009}

Several approaches are currently used for calculating rate constants and KIE's
in situations where quantum effects are not negligible. One approach consists
in adding a tunneling correction to transition state
theory,\cite{Ramazani:2013} others use an approximation for the propagator by
treating it semiclassically\cite{Olsson_Warshel:2004}
or by treating only one or two degrees of freedom quantum
mechanically.\cite{Banks_Clary:2007} Another promising method is the ring
polymer molecular dynamics\cite{Allen_Suleimanov:2013} (RPMD), which can
partially capture both quantum effects and classical recrossing. Finally,
there are various quantum generalizations of the transition state theory.
Among these so-called quantum transition state
theories\cite{Hansen_Andersen:1994,Pollak:2012} belongs the quantum instanton
(QI) approximation to the rate constant,\cite{Miller_Yang:2003} i.e., the
method whose efficiency we attempt to increase in the present paper. The
QI\ approximation is motivated by the semiclassical instanton theory
\cite{Miller:1974,Chapman_Miller:1975,Miller_Tromp:1983,Ceotto:2012} and, as
the name suggests, takes into account only the zero-time properties of the
reactive flux-flux correlation function; however, in contrast to the
semiclassical instanton, the QI approximation treats the Boltzmann operator
exactly quantum mechanically. This improvement makes QI quite accurate as
verified in many previous applications of the
method.\cite{Zhao_Miller:2004,Ceotto_Miller:2004,Wang_Zhao:2007,Wang_Zhao:2009,Wang_Zhao:2011,Wang_Zhao:2012}%

QI theory expresses the reaction rate in terms of imaginary-time correlation
functions, which, in turn, can be evaluated by path integral (PI) Monte Carlo (MC)
methods.\cite{Yamamoto_Miller:2004} As for KIE's, the problem can be
simplified further by using thermodynamic integration.\cite{Vanicek_Aoiz:2005}
The resulting method, however, has a drawback common to all PI methods: it
operates in a configuration space of greatly increased dimensionality, leading
to high computational cost. Indeed, the quantum limit is approached as the
number of dimensions goes to infinity. Several approaches have been proposed
to bypass the problem and this paper combines two of them to accelerate the QI calculations.

The first approach employs Boltzmann operator factorizations of higher order
of accuracy. The resulting PI representations of relevant
quantities exhibit faster convergence to the quantum limit, allowing to reduce
the dimensionality of the
calculation.\cite{Jang_Voth:2001,Yamamoto:2005,Perez_Tuckerman:2011,Buchowiecki_Vanicek:2013,Marsalek_Tuckerman:2014,Engel_Major:2015} The
second approach uses improved estimators with lower statistical errors, which
permit shortening the MC
simulation.\cite{Yang_Miller:2006,Vanicek_Miller:2007} In this work we combine
these two strategies, and, in addition, propose several new estimators. We
then test the resulting method on two systems: the model $\mathrm{\cdot
H_{\alpha}+H_{\beta}H_{\gamma}\rightarrow H_{\alpha}H_{\beta}+\cdot H_{\gamma
}}$ rearrangement, for which we also evaluate the resulting gain in
computational efficiency, and the reaction $\mathrm{CH_{4}+\cdot
H\rightleftharpoons\cdot CH_{3}+H_{2}}$, a process whose KIE's were studied in
detail both experimentally and theoretically, with classical TST, several of
its corrected versions,\cite{Pu_Truhlar:2002,Schatz_Dunning:1984} reduced
dimensionality quantum dynamics,\cite{Kerkeni_Clary:2004} and
RPMD.\cite{Li_Guo:2013}

The rest of the paper is organized as follows: After outlining
the derivation of the QI approximation for the rate constant in Sec.~\ref{sec:QI_formalism}, in the central
Sec.~\ref{sec:PI_implementation} we first show how this approximation can be
combined with the PI formalism and then explain in detail the two strategies
to improve numerical performance of the standard PI implementation. The
numerical results are presented in Sec.~\ref{sec:Application}, while
Sec.~\ref{sec:conclusions} concludes the paper. To facilitate the reading,
our notation is summarized in Table \ref{tab:System_notation}.

\begin{table}
\caption{Summary of the notation used in this paper for a system of $ N $ particles in a $ D$-dimensional Euclidean space.
$ m_{i}$ is the mass of particle $ i$; $ \mathbf{v}$ and $ \mathbf{w}$ are vectors defined in
the $ ND$-dimensional configuration space, while $ \mathrm{v}_{i}$ and $ \mathrm{w}_{i}$
are their $ D$-dimensional components corresponding to particle $ i$; $ \mathbf{A}$ is a Hermitian matrix
defined over the configuration space, and $ \mathrm{A}_{ij}$ is its $ D\times D$ dimensional submatrix
containing only the columns corresponding
to particle $i$ and rows corresponding to particle $j$.
\hfill}\label{tab:System_notation}
\begin{ruledtabular}
\begin{tabular}{cm{0.5\linewidth}}
Expression & Comment \tabularnewline
\hline
\hline
 $ \nabla_{i}$ & gradient with respect to coordinates of particle $ i $ \tabularnewline
\hline
 $ \mathrm{v}_{i}\cdot\mathrm{w}_{i}:=\sum_{j=1}^{D}\mathrm{v}_{i,j}%
\cdot\mathrm{w}_{i,j}$ & standard dot product of $ \mathrm{v}_{i}$ and $ \mathrm{w}_{i}$ in the $ D$-dimensional Euclidean space \tabularnewline
\hline
 $\langle\mathbf{v},\mathbf{w}\rangle_{s}:=\sum_{i=1}^{N}(m_{i})^{s}\mathrm{v}_{i}\cdot\mathrm{w}_{i} $ &
mass-weighted inner product of $ \mathbf{v}$ and $ \mathbf{w}$ in the system's configuration space, where $s\in\{-1,~0,~1\}$
on the right-hand side, while on the left-hand side a corresponding shorthand notation $s\in\{-,~0,~+\}$ is used\tabularnewline
\hline
 $ ||\mathbf{v}||_{s}:=\sqrt{\langle\mathbf{v},\mathbf{v}\rangle_{s}} $ & mass-weighted norm of a configuration space vector\tabularnewline
\hline
$ \langle\mathbf{v},\mathbf{A},\mathbf{w}\rangle_{su}:=\sum_{i=1}^{N}\sum_{j=1}^{N}%
 m_{i}^{s}m_{j}^{u}\mathrm{v}_{i}\cdot\mathrm{A}_{ij}\cdot\mathrm{w}_{j} $
 & matrix product of $ \mathbf{A}$ with $ \mathbf{v}$ and $ \mathbf{w}$; 
 the same shorthand notation is used for $ s$ and $ u $
 as in $\langle\mathbf{v},\mathbf{w}\rangle_{s}$\tabularnewline
\end{tabular}
\end{ruledtabular}

\end{table}

\section{Quantum instanton formalism\label{sec:QI_formalism}}

The QI approximation for the thermal rate constant $k(T)$ can be derived from
the exact Miller-Schwartz-Tromp formula,\cite{Miller_Tromp:1983}
\begin{equation}
k(T)\,Q_{r}=\int_{0}^{\infty}C_{\text{ff}}(t)\mathit{d}t\text{,}
\label{eq:MST}%
\end{equation}
expressing the product of the rate constant with the reaction partition
function $Q_{r}$ as the time integral of the flux-flux correlation function%
\begin{equation}
C_{\text{ff}}(t):=C_{\hat{F}_{a}\hat{F}_{b}}(t)\text{,}%
\end{equation}
where
\begin{equation}
 C_{\hat{A}\hat{B}}(t):=\mathrm{Tr}\left(\hat{A}e^{-(\beta/2-\mathrm{i}t)\hat{H}}
 \hat{B}e^{-(\beta/2+\mathrm{i}t)\hat{H}}\right)
\end{equation}
is the symmetrized correlation function of operators $\hat{A}$ and $\hat{B}$ at
temperature $T=1/(k_{B}\beta)$ and time $ t$,
\begin{equation}
\hat{F}_{\gamma}:=-\frac{\mathrm{i}}{\hbar}[h[\xi_{\gamma}(\hat{\mathbf{r}%
})],\hat{H}]
\end{equation}
is the operator of flux through dividing surface (DS) $\ \gamma\in\{a,b\}$, $ \mathbf{r}$ is
the position vector in the $ ND$-dimensional configuration space ($ N$ is the number of atoms,
$ D$ is the number of spatial dimensions), and $h$
is the Heaviside function [i.e., $h(x)=1$ for $x\geq0$ and $h(x)=0$ for
$x<0$]. The two DS's $a$ and $b$ completely separate the reactant
and product regions, and are defined by the equation $\xi_{\gamma}%
(\mathbf{r})=0$. In addition, $\xi_{\gamma}$ are chosen so that $\xi_{\gamma
}(\mathbf{r})>0$ for $\mathbf{r}$ in the product region and $\xi_{\gamma
}(\mathbf{r})<0$ in the reactant region.

The QI approximation can be derived by applying the steepest
descent approximation to Eq.~\eqref{eq:MST}.
\cite{Vanicek_Aoiz:2005,Ceotto_Miller_private_communication:2003} First, one
multiplies and divides the integrand of Eq. (\ref{eq:MST}) by the so-called
delta-delta correlation function%
\begin{equation}
C_{\text{dd}}(t):=C_{\hat{\Delta}_{a}\hat{\Delta}_{b}}(t)\text{,}
\label{eq:QI}%
\end{equation}
where $\Delta_{\gamma}$ is the normalized delta function%
\begin{equation}
\Delta_{\gamma}(\mathbf{r})=\delta\lbrack\xi_{\gamma}(\mathbf{r})]||\nabla
\xi_{\gamma}(\mathbf{r})||_{-} \label{eq:normalised_delta}%
\end{equation}
and $||\cdot||_{-}$ is the norm of a covariant vector (see Table~\ref{tab:System_notation}).
Then one assumes that $C_{\text{ff}}(t)$ decays sufficiently fast so that the
main contribution to the integral in Eq.~\eqref{eq:MST} comes from $t$ close
to zero (hence the name \textquotedblleft quantum instanton\textquotedblright),
and that for these short times the ratio $C_{\text{ff}}(t)/C_{\text{dd}%
}(t)$ remains approximately constant and given by $C_{\text{ff}}%
(0)/C_{\text{dd}}(0)$. One can therefore evaluate the time integral in
Eq.~(\ref{eq:MST}) with the steepest descent approximation,%

\begin{align}
\int_{0}^{\infty}C_{\text{ff}}(t)\mathit{d}t  &  =\int_{0}^{\infty}%
\frac{C_{\text{ff}}(t)}{C_{\text{dd}}(t)}C_{\text{dd}}(t)\mathit{d}%
t\nonumber\\
&  \approx\frac{C_{\text{ff}}(0)}{C_{\text{dd}}(0)}\int_{0}^{\infty
}C_{\text{dd}}(t)\mathit{d}t\nonumber\\
&  \approx\frac{C_{\text{ff}}(0)}{C_{\text{dd}}(0)}\frac{\hbar\sqrt{\pi}}%
{2}\frac{C_{\text{dd}}(0)}{\Delta H}, \label{eq:MST_pre_QI}%
\end{align}
obtaining the QI expression for the rate constant%
\begin{equation}
k_{\mathrm{QI}}=\frac{\hbar\sqrt{\pi}}{2}\frac{C_{\text{dd}}(0)}{Q_{r}}%
\frac{C_{\text{ff}}(0)/C_{\text{dd}}(0)}{\Delta H}\text{,} \label{eq:QI_basic}%
\end{equation}
where
\begin{equation}
\Delta H=\hbar\sqrt{-\frac{\ddot{C}_{\text{dd}}(0)}{2C_{\text{dd}}(0)}}
\label{eq:dH_definition}%
\end{equation}
is a certain energy variance. For reasons that will become clear below, we
keep $C_{\text{dd}}(0)$ in Eq.~\eqref{eq:QI_basic}, even though it may seem to
cancel out.

The last question to be addressed is how to choose positions of the optimal DS's.
From semiclassical considerations it follows that the best
choice is to require that $C_{\text{dd}}(0)$ be a saddle point with respect to
$\xi_{a}$ and $\xi_{b}$;\cite{Miller_Yang:2003} if $\xi_{\gamma}$ are
controlled by a set of parameters $\{\eta_{k}^{(\gamma)}\},$ the stationarity
condition becomes%
\begin{equation}
\frac{\partial C_{\text{dd}}}{\partial\eta_{k}^{(\gamma)}}=0\mathrm{.}
\label{eq:stationarity}%
\end{equation}

\section{General path integral implementation\label{sec:PI_implementation}}

The QI approximation allows expressing the rate constant in terms of the
reactant partition function and properties of flux-flux and delta-delta
correlation functions at time $t=0$. In this section, we first explain how the
PI formalism allows transforming the quantum problem of finding these
quantities to a classical one, applied to the so-called polymer
chain,\cite{Feynman_Hibbs:1965, Chandler_Wolynes:1981} and then describe an
efficient implementation allowing a significant acceleration of calculations
of the KIE's.

One of our goals is using higher-order factorizations of the Boltzmann
operator in order to accelerate the convergence of the KIE's to the quantum
limit. In Subsec.~\ref{subsec:Boltzmann_splittings}, we therefore present a
general derivation of the PI expression for the Boltzmann operator matrix
element, valid for all Boltzmann operator factorizations used in this work,
and in Subsec.~\ref{subsec:PI_Qr_Cdd} we obtain general PI expressions for
$Q_{r}$ and $C_{\mathrm{dd}}(0)$. In Subsecs.~\ref{subsec:Estimators_dH_Cff}
and~\ref{subsec:Thermodynamic_integration}, we explain how all quantities
necessary for computing the KIE within the QI approximation can be expressed
in terms of thermodynamic averages over ensembles corresponding to PI
expressions for $Q_{r}$ and $C_{\mathrm{dd}}(0)$; in
Subsec.~\ref{subsec:Centroid-virial-estimators} we derive estimators allowing
to calculate these averages with a lower statistical error and therefore
significantly accelerating statistical convergence, which is our second main
goal. Some of the more tedious derivations are deferred to the Appendix.

\subsection{Lie-Trotter, Takahashi-Imada, and Suzuki factorizations of the
imaginary-time path integral propagator \label{subsec:Boltzmann_splittings}}

The coordinate matrix element of the Boltzmann operator at temperature
$T=1/(k_{B}\beta)$ can be rewritten as a matrix element of the product of
$P\in%
\mathbb{N}
$ Boltzmann operators at a higher temperature inversely proportional to the
parameter $\epsilon:=\beta/P$:%
\begin{equation}
\langle\mathbf{r}^{(a)}|e^{-\beta\hat{H}}|\mathbf{r}^{(b)}\rangle
=\langle\mathbf{r}^{(a)}|(e^{-\epsilon\hat{H}})^{P}|\mathbf{r}^{(b)}%
\rangle\text{.} \label{eq:Boltzmann_el_rewritten}%
\end{equation}
We next consider three possible high-temperature factorizations of the
Boltzmann operator:

1. The symmetrized version of the Lie-Trotter factorization:%
\begin{equation}
e^{-\epsilon\hat{H}}=e^{-\epsilon\hat{V}/2}e^{-\epsilon\hat{T}}e^{-\epsilon
\hat{V}/2}+O\left(  \epsilon^{3}\right)  \text{.} \label{eq:Lie_Trotter}%
\end{equation}
This second-order factorization, which we will for simplicity denote by LT is the one most commonly used for discretizing
the imaginary-time Feynman PI.

2. The Takahashi-Imada (TI) factorization:\cite{Takahashi_Imada:1984}%
\begin{equation}
\mathrm{Tr}\left(  e^{-\epsilon\hat{H}}\right)  =\mathrm{Tr}\left(
e^{-\epsilon\hat{V}_{\mathrm{TI}}/2}e^{-\epsilon\hat{T}}e^{-\epsilon\hat
{V}_{\mathrm{TI}}/2}\right)  +O\left(  \epsilon^{5}\right)  , \label{eq:TI}%
\end{equation}
where%
\begin{equation}
\hat{V}_{\mathrm{TI}}:=\hat{V}+\frac{1}{24}\epsilon^{2}[\hat{V},[\hat{T}%
,\hat{V}]] \label{eq:TI_V_eff}%
\end{equation}
is an effective one-particle potential. This fourth-order factorization significantly
accelerates the convergence to the quantum limit of the PI expression for
$Q_{r}$. However, it only behaves as a fourth-order factorization when it is
used for evaluating the trace of the Boltzmann operator. If one naively
removes the $\operatorname{Tr}$ in Eq.~(\ref{eq:TI}), and applies the
resulting factorization
\begin{equation}
e^{-\epsilon\hat{H}}\approx e^{-\epsilon\hat{V}_{\mathrm{TI}}/2}%
e^{-\epsilon\hat{T}}e^{-\epsilon\hat{V}_{\mathrm{TI}}/2} \label{eq:TI_bad}%
\end{equation}
to off-diagonal elements, which are required for PI representations of
$C_{\text{dd}}(0)$ and $C_{\text{ff}}(0)$, one obtains only second-order
convergence, and no numerical advantage over the LT factorization.
Since it will allow us to provide a single derivation of many quantities for
different factorizations, we will abuse terminology and refer to
Eq.~(\ref{eq:TI_bad}) also as \textquotedblleft
Takahashi-Imada\textquotedblright\ factorization,\ keeping in mind that the
original authors were aware that their splitting is of the fourth order only in
the context of Eq.~(\ref{eq:TI}).

3. The fourth-order Suzuki-Chin (SC) factorization (Ref. \onlinecite{Chin:1997}, motivated by Ref. \onlinecite{Suzuki:1995}):
\begin{equation}
e^{-\epsilon\hat{H}}=e^{-\epsilon\hat{V}_{\mathrm{e}}/6}e^{-\epsilon\hat{T}%
/2}e^{-2\epsilon\hat{V}_{\mathrm{m}}/3}e^{-\epsilon\hat{T}/2}e^{-\epsilon
\hat{V}_{\mathrm{e}}/6}+O\left(  \epsilon^{5}\right),
\end{equation}
where
\begin{align}
\hat{V}_{\mathrm{e}}  &  :=\hat{V}+\frac{\alpha}{6}\epsilon^{2}[\hat{V}%
,[\hat{T},\hat{V}]]\text{ and}\label{eq:Suzuki_V_eff_1}\\
\hat{V}_{\mathrm{m}}  &  :=\hat{V}+\frac{(1-\alpha)}{12}\epsilon^{2}[\hat
{V},[\hat{T},\hat{V}]] \label{eq:Suzuki_V_eff_2}%
\end{align}
are the \textquotedblleft endpoint\textquotedblright\ and \textquotedblleft
midpoint\textquotedblright\ effective one-particle potentials. The
dimensionless parameter $\alpha$ can assume an arbitrary value, but evidence
in the literature\cite{Jang_Voth:2001,Perez_Tuckerman:2011} suggests that
$\alpha=0$ gives superior results in most PI simulations, and hence it was
also the value used in our calculations.

Now we use one of the three PI splittings for each of the $P$ high-temperature
factors in Eq.~\eqref{eq:Boltzmann_el_rewritten}, with the caveat that for the
SC factorization (only) we replace $P$ with $P/2$ (so $P$ must be even)
and $\epsilon=\beta/P$ with $\epsilon=2\beta/P$ in
Eq.~\eqref{eq:Boltzmann_el_rewritten}. After inserting $(P-1)$ resolutions of
identity in the coordinate basis in front of every kinetic factor (except the
first one), we obtain
\begin{equation}
\langle\mathbf{r}^{(a)}|e^{-\beta\hat{H}}|\mathbf{r}^{(b)}\rangle
=\lim_{P\rightarrow\infty}C\int\mathit{d}\mathbf{r}^{(1)}\cdot\cdot
\cdot\mathit{d}\mathbf{r}^{(P-1)}\mathrm{exp}\left[  -\beta\tilde{\Phi
}(\mathbf{r}^{(a)},\mathbf{r}^{(1)},...,\mathbf{r}^{(P-1)},\mathbf{r}%
^{(b)})\right]  \text{,} \label{eq:Boltzmann_matrix_element}%
\end{equation}
where the effective potential $\tilde{\Phi}$ and prefactor $C$ are defined as
\begin{align}
\tilde{\Phi}  &  :=\frac{P}{2\hbar^{2}\beta^{2}}\sum_{s=1}^{P}||\mathbf{r}%
^{(s)}-\mathbf{r}^{(s-1)}||_{+}^{2}+\frac{1}{P}\sum_{s=0}^{P}\tilde{w}%
_{s}V_{\mathrm{eff}}^{(s)}(\mathbf{r}^{(s)}),\label{eq:tilde_Phi}\\
C  &  :=\left(  \frac{P}{2\hbar^{2}\pi\beta}\right)  ^{NDP/2}\left(
\prod_{i=1}^{N}m_{i}\right)^{DP/2}.
\end{align}
In the expression for $\tilde{\Phi}$, we use the notation $\mathbf{r}%
^{(P)}:=\mathbf{r}^{(b)}$, $\mathbf{r}^{(0)}:=\mathbf{r}^{(a)}$ for the
boundary points; $ N$ is the number of atoms, $ D$ is the number of spatial dimensions,
$m_{i}$ is the mass of particle $ i$, $||\cdot||_{+}$ is the norm of a contravariant vector (see Table~\ref{tab:System_notation}),
and $V_{\mathrm{eff}}^{(s)}$ is the effective one-particle
potential,
\begin{equation}
V_{\mathrm{eff}}^{(s)}:=V+\left(  \frac{\beta}{P}\right)  ^{2}d_{s}%
V_{\mathrm{grad}},
\end{equation}
where%
\begin{equation}
V_{\mathrm{grad}}(\mathbf{r})=\hbar^{2}||\nabla V(\mathbf{r})||_{-}^{2}%
\end{equation}
is the coordinate representation of the commutator term in
Eqs.~\eqref{eq:TI_V_eff}, \eqref{eq:Suzuki_V_eff_1}, and~\eqref{eq:Suzuki_V_eff_2}. 
In the context of discretized PI's, the integer $ P$ is often referred to as the Trotter number.

The coefficient $d_{s}$ for the fourth-order correction of an effective
one-particle potential depends on the splitting used:%
\begin{equation}
d_{s}=%
\begin{cases}
0, & \text{LT splitting,}\\
1/24, & \text{TI splitting,}\\
\alpha/6, & \text{SC splitting and }s\text{ even,}\\
(1-\alpha)/12, & \text{SC splitting and }s\text{ odd.}%
\end{cases}
\end{equation}
The weights $\tilde{w}_{s}$ in the sum over effective one-particle potentials
also depend on the splitting: for endpoint $s$ (i.e., $s=0,P$) these weights
are $\tilde{w}_{s}=1/2$ for the LT and TI splittings,
and $\tilde{w}_{s}=1/3$ for the SC splitting; for other values of $s$,
$\tilde{w}_{s}=1$ for the LT and TI splittings, whereas
for the SC splitting, $\tilde{w}_{s}=4/3$ for odd $s$ and $\tilde{w}%
_{s}=2/3$ for even $s$. Expression (\ref{eq:Boltzmann_matrix_element})
becomes exact as $P$ goes to infinity.

\subsection{Path integral representation of the partition function and
delta-delta correlation function \label{subsec:PI_Qr_Cdd}}

From Eq.~\eqref{eq:Boltzmann_matrix_element} it is straightforward to obtain
the PI representation $Q_{r,P}$ of the reactant partition function $Q_{r}$; in
particular,
\begin{align}
Q_{r}  &  =\mathrm{Tr}\left(  e^{-\beta\hat{H}}\right)  =\int\langle
\mathbf{r}|e^{-\beta\hat{H}}|\mathbf{r}\rangle\mathit{d}\mathbf{r}%
=\lim_{P\rightarrow\infty}Q_{r,P}\text{,}\\
Q_{r,P}  &  =\int\mathit{d}\{\mathbf{r}^{(s)}\}\rho_{r}(\{\mathbf{r}%
^{(s)}\})\text{,}\label{eq:Qr_PI}\\
\rho_{r}(\{\mathbf{r}^{(s)}\})  &  =C \exp\left[  -\beta\Phi(\{\mathbf{r}%
^{(s)}\})\right]  \text{.}%
\end{align}
(In general, we will distinguish between a quantity $A$ and its PI
representation $A_{P}$ for a given value of $P$ by adding an additional
subscript $P$.) By $\int\mathit{d}\{\mathbf{r}^{(s)}\}$ we mean integration
over all $\mathbf{r}^{(s)}$, $s\in\{1,...,P\}$; $\rho_{r}(\{\mathbf{r}%
^{(s)}\})$ can be regarded as the thermal distribution of $\{\mathbf{r}%
^{(s)}\}$ of the new system; $\Phi$ is the closed-loop version of $\tilde
{\Phi}$, i.e.,
\begin{equation}
\Phi(\{\mathbf{r}^{(s)}\}):=\tilde{\Phi}(\mathbf{r}^{(P)},\mathbf{r}%
^{(1)},...,\mathbf{r}^{(P)})=\frac{P}{2\hbar^{2}\beta^{2}}\sum_{s=1}%
^{P}||\mathbf{r}^{(s)}-\mathbf{r}^{(s-1)}||_{+}^{2}+\frac{1}{P}\sum_{s=1}%
^{P}w_{s}V_{\mathrm{eff}}^{(s)}(\mathbf{r}^{(s)})\text{.}%
\end{equation}
From now on we will always consider closed loops such that $\mathbf{r}%
^{(0)}=\mathbf{r}^{(P)}$. The difference between the new weights $w_{s}$ and
the old weights $\tilde{w}_{s}$ is that $w_{P}=1$ for the LT or TI
splittings and $w_{P}=2/3$ for the SC splitting, for which
we also require $P$ to be even.

We can now see that the PI representation of the quantum partition function
$Q_{r}$ is identical to the classical partition function $Q_{r,\text{cl}}$ of
a system (called \textquotedblleft polymer chain\textquotedblright) where
every original particle is replaced with $P$ pseudoparticles connected by
harmonic forces. Also note that for $P=1$ and the LT factorization
our PI expression reduces to the expression for the classical Boltzmann distribution.

For $C_{\text{dd}}(0)$ we have, analogously,
\begin{align}
C_{\mathrm{dd},P}  &  =\int\mathit{d}\{\mathbf{r}^{(s)}\}\rho^{\ddagger
}(\{\mathbf{r}^{(s)}\})\text{,}\label{eq:C_dd(PI)}\\
\rho^{\ddagger}(\{\mathbf{r}^{(s)}\})  &  =C\Delta_{a}(\mathbf{r}%
^{(P/2)})\Delta_{b}(\mathbf{r}^{(P)})\exp\left[  -\beta\Phi(\{\mathbf{r}%
^{(s)}\})\right]  \text{.}%
\end{align}
(we shall omit the time argument of $C_{\text{dd}}$ and $C_{\text{ff}}$ if it
equals 0). Note that $\rho^{\ddagger}(\{\mathbf{r}^{(s)}\})$ differs from $\rho_{r}
(\{\mathbf{r}^{(s)}\})$ by the two delta constraints imposed on $\mathbf{r}%
^{(P/2)}$ and $\mathbf{r}^{(P)}$.

In the rest of the section we will show how the QI expression for the KIE can be
rewritten in terms of classical thermodynamic averages over $\rho_{r}$ and
$\rho^{\ddagger}$. Expressions for the corresponding estimators will be
presented in a general way valid for all Boltzmann operator splittings
considered in this work and as such will contain the main part common for all
splittings and a part which corresponds to the fourth-order corrections and is
only non-zero if a splitting other than LT is used; 
since this additional part depends on the gradient of the potential energy we 
will denote it by adding ``grad'' subscript to the name of the estimator.
Although it is one of the main results of this work, for clarity the derivation of the parts
associated with the fourth-order factorizations will be left for Appendix A.
Before we proceed it is necessary to point out relative costs of running MC
simulations over $\rho_{r}$ and $\rho^{\ddagger}$ obtained with different
Boltzmann operator splittings. While the use of the LT splitting only requires
potential energy for each $\mathbf{r}^{(s)}$, the SC splitting
with $0<\alpha<1$ and the TI factorization also require
the gradient of energy for each $\mathbf{r}^{(s)}$, and the SC splittings
with $\alpha=0$ and $\alpha=1$ require gradients for $\mathbf{r}^{(s)}$
with $s$ odd and even, respectively.

\subsection{Estimators for constrained quantities
\label{subsec:Estimators_dH_Cff}}

Within the PI formalism both the energy spread $\Delta H$ and the flux factor
$C_{\mathrm{ff}}/C_{\mathrm{dd}}$ can be expressed as thermodynamic averages
over the ensemble whose configurations are weighted by $\rho^{\ddagger
}(\{\mathbf{r}^{(s)}\})$.\cite{Yamamoto_Miller:2004} In order to obtain the PI
representation of $\ddot{C}_{\text{dd}}(0)$ and $\Delta H^{2}$, it is
convenient to perform the Wick rotation and define a new function
\begin{equation}
\overline{C}_{\text{dd}}(\zeta):=C_{\text{dd}}\left(  -\frac{\mathrm{i}%
\zeta\hbar}{2}\right)
\end{equation}
of a complex argument $\zeta$. Supposing that $C_{\text{dd}}(t)$ is analytic,%
\begin{equation}
\ddot{C}_{\text{dd}}(t)=-\frac{4}{\hbar^{2}}\left.\frac{\partial^{2}}{\partial
\zeta^{2}}\overline{C}_{\text{dd}}\left(\zeta\right)\right|_{\zeta=2\mathrm{i}t/\hbar}\text{.} \label{C_dd_complex}%
\end{equation}
The PI representation of $\overline{C}_{\text{dd}}(\zeta)$ is%
\begin{equation}
\overline{C}_{\text{dd},P}(\zeta)=\overline{C}\int\mathit{d}\{\mathbf{r}%
^{(s)}\}\Delta_{a}(\mathbf{r}^{(P/2)})\Delta_{b}(\mathbf{r}^{(P)})\exp\left(
-\beta^{+}\tilde{\Phi}^{+}-\beta^{-}\tilde{\Phi}^{-}\right)  \text{,}
\label{eq:C_dd_zeta}%
\end{equation}
with
\begin{align}
\beta^{+}  &  =\beta+\zeta\text{,}\\
\beta^{-}  &  =\beta-\zeta\text{,}%
\end{align}
prefactor
\begin{equation}
\overline{C}=\left(  \frac{P}{2\hbar^{2}\pi\sqrt{\beta^{2}-\zeta^{2}}}\right)
^{NDP/2}\left(  \prod_{i=1}^{N}m_{i}\right)  ^{DP/2}\text{,}%
\end{equation}
and two ``partial'' effective potentials
\begin{align}
\tilde{\Phi}^{+}  &  =\frac{P}{2\hbar^{2}(\beta^{+})^{2}}\sum_{s=1}%
^{P/2}||\mathbf{r}^{(s)}-\mathbf{r}^{(s-1)}||_{+}^{2}+\frac{1}{P}\sum
_{s=0}^{P/2}\tilde{\tilde{w}}_{s}V_{\mathrm{eff}}^{(s)}(\mathbf{r}%
^{(s)})\text{,}\label{eq:Phi_plus}\\
\tilde{\Phi}^{-}  &  =\frac{P}{2\hbar^{2}(\beta^{-})^{2}}\sum_{s=P/2+1}%
^{P}||\mathbf{r}^{(s)}-\mathbf{r}^{(s-1)}||_{+}^{2}+\frac{1}{P}\sum
_{s=P/2}^{P}\tilde{\tilde{w}}_{s}V_{\mathrm{eff}}^{(s)}(\mathbf{r}%
^{(s)})\text{,} \label{eq:Phi_minus}%
\end{align}
where $\tilde{\tilde{w}}_{s}=\tilde{w}_{s}$ for all $ s $ except for $s=P/2$,
for which $\tilde{\tilde{w}}_{P/2}=\tilde{w}_{P}=\tilde{w}_{0}$. 
The effective potentials $ \tilde{\Phi}^{+}$ and $ \tilde{\Phi}^{-}$ in Eq. (\ref{eq:C_dd_zeta})
are obtained in a similar manner as 
$\tilde{\Phi}$ was obtained in Eq. (\ref{eq:Boltzmann_matrix_element}).
The difference is that instead of the matrix element of the Boltzmann operator $ \exp(-\beta\hat{H})$ 
one considers an element of $ \exp(-\beta^{+}\hat{H}/2)$ or $ \exp(-\beta^{-}\hat{H}/2)$,
and the exponential operators are discretized into $ P/2$ rather than $ P$ parts.
As a result, expressions (\ref{eq:Phi_plus})-(\ref{eq:Phi_minus}) for $\tilde{\Phi}^{+}$
and $ \tilde{\Phi}^{-}$ can be obtained from the one for $ \tilde{\Phi}$ [Eq. (\ref{eq:tilde_Phi})] if 
$ \beta$ is replaced with $ \beta^{+}/2$ and $ \beta^{-}/2$, respectively, and $ P $ is replaced with $ P/2$.
After differentiating expression~(\ref{eq:C_dd_zeta}) with respect to $\zeta$, using
Eq.~(\ref{C_dd_complex}) to go from $\overline{C}_{\text{dd},P}^{\prime\prime
}(\zeta)$ back to $\ddot{C}_{\text{dd}}(t)$, and noting that $\mathit{d}%
\zeta=\mathit{d}\beta^{+}=-\mathit{d}\beta^{-}$, one obtains%
\begin{equation}
\ddot{C}_{\text{dd},P}(0)=-\frac{1}{\hbar^{2}}C\int\mathit{d}\{\mathbf{r}%
^{(s)}\}\left(G+F^{2}\right)  \rho^{\ddagger}(\{\mathbf{r}^{(s)}\})\text{,}
\label{eq:C_dd_dotdot}%
\end{equation}
with
\begin{align}
G  &  =4\left[  \frac{d^{2}\mathrm{ln}C}{\mathit{d}\beta^{2}}-\frac
{\mathit{d}^{2}(\beta^{+}\tilde{\Phi}^{+})}{\mathit{d}(\beta^{+})^{2}}%
-\frac{\mathit{d}^{2}(\beta^{-}\tilde{\Phi}^{-})}{\mathit{d}(\beta^{-})^{2}%
}\right]  \text{,}\label{eq:G_factor}\\
F  &  =2\left[  \frac{\mathit{d}(\beta^{+}\tilde{\Phi}^{+})}{\mathit{d}%
\beta^{+}}-\frac{\mathit{d}(\beta^{-}\tilde{\Phi}^{-})}{\mathit{d}\beta^{-}%
}\right]  \text{.} \label{eq:F_factor}%
\end{align}
After the substitution of expressions (\ref{eq:C_dd(PI)}) and (\ref{eq:C_dd_dotdot}) for
$C_{\mathrm{dd}, P}$ and $\ddot{C}_{\mathrm{dd}, P}$ into the definition
(\ref{eq:dH_definition}) of $\Delta H^{2}$,
the estimator for $ \Delta H^{2}$ takes the form
\begin{equation}
(\Delta H^{2})_{P,\mathrm{est}}=\frac{G+F^{2}}{2}%
\end{equation}
if $\rho^{\ddagger}(\{\mathbf{r}^{(s)}\})$ is used as the weight function.
From now on, if a quantity $ A$ can be expressed as a classical thermodynamic average,
we will denote the corresponding estimator by $A_{\mathrm{est}}$ (the density
function over which the averaging is performed will not be denoted explicitly since
this will always be clear from the context).

Explicit differentiation in Eqs. (\ref{eq:G_factor}) and (\ref{eq:F_factor})
leads to the so-called thermodynamic estimator,\cite{Yamamoto_Miller:2004}%
\begin{align}
G_{\mathrm{th}}  &  =\frac{2NDP}{\beta^{2}}-\frac{4P}{\hbar^{2}\beta^{3}}%
\sum_{s=1}^{P}||\mathbf{r}^{(s)}-\mathbf{r}^{(s-1)}||_{+}^{2}+G_{\mathrm{th,grad}}\text{,}\label{eq:G_th}\\
F_{\mathrm{th}}  &  =\frac{2}{P}\left(  \sum_{s=1}^{P/2-1}-\sum_{s=P/2+1}%
^{P-1}\right)  w_{s}V_{\mathrm{eff}}^{(s)}(\mathbf{r}^{(s)})-\frac{P}%
{\hbar^{2}\beta^{2}}\left(  \sum_{s=1}^{P/2}-\sum_{s=P/2+1}^{P}\right)
||\mathbf{r}^{(s)}-\mathbf{r}^{(s-1)}||_{+}^{2}+F_{\mathrm{grad}}\text{.}
\label{eq:F_th}%
\end{align}

The ratio $C_{\text{ff}}/C_{\text{dd}}$ can be computed by the Metropolis
algorithm as well. To obtain the corresponding estimator we first note that
the flux operator can be expressed as
\begin{equation}
\hat{F}_{\gamma}=\frac{1}{2}\left\{  \delta\lbrack\xi_{\gamma}(\hat
{\mathbf{r}})]\langle\nabla\xi_{\gamma}(\hat{\mathbf{r}}),\hat{\mathbf{p}%
}\rangle_{-}+\langle\nabla\xi_{\gamma}(\hat{\mathbf{r}}),\hat{\mathbf{p}%
}\rangle_{-}\delta\lbrack\xi_{\gamma}(\hat{\mathbf{r}})]\right\}  \text{.}%
\end{equation}
Combining $\hat{F}_{\gamma}$ with the PI representation of the Boltzmann
operator, one obtains\cite{Yamamoto_Miller:2004}
\begin{equation}
C_{\text{ff},P}=C\int\mathit{d}\{\mathbf{r}^{(s)}\}f_{\mathrm{v}}%
\rho^{\ddagger}(\{\mathbf{r}^{(s)}\})\text{,} \label{eq:C_ff}%
\end{equation}
where $f_{\mathrm{v}}$ is the so-called velocity factor,%
\begin{equation}%
\begin{split}
f_{\mathrm{v}}=  &  \frac{\hbar^{2}}{4}\left\{  \beta^{2}\prod_{\gamma
=a,b}\left\langle \nabla\xi_{\gamma}(\mathbf{r}_{\gamma}),\left(
\frac{\partial\tilde{\Phi}^{+}}{\partial\mathbf{r}_{\gamma}}-\frac
{\partial\tilde{\Phi}^{-}}{\partial\mathbf{r}_{\gamma}}\right)  \right\rangle
_{-}\right. \\
&  -\left.
\vphantom{\beta^{2}\prod_{\gamma=a,b}\left\langle\nabla\xi_{\gamma} (\mathbf{r}_{\gamma}) \cdot\left(\frac{\partial\tilde{\Phi}^{+}}{\partial\mathbf{r}_{\gamma}}- \frac{\partial\tilde{\Phi}^{-}}{\partial\mathbf{r}_{\gamma}}\right)\right\rangle_{-}}\beta
\left\langle \nabla\xi_{a}(\mathbf{r}^{(P/2)}),\frac{\partial^{2}(\tilde{\Phi
}^{+}+\tilde{\Phi}^{-})}{\partial\mathbf{r}^{(P/2)}\partial\mathbf{r}^{(P)}%
},\nabla\xi_{b}(\mathbf{r}^{(P)})\right\rangle_{--}\right\} \\
&  /\{\prod_{\gamma=a,b}||\nabla\xi_{\gamma}(\mathbf{r}_{\gamma}%
)||_{-}\}\text{,}%
\end{split}
\label{eq:f_general}%
\end{equation}
$\mathbf{r}_{a}=\mathbf{r}^{(P/2)}$, $\mathbf{r}_{b}=\mathbf{r}^{(P)}$,
$\langle\cdot,\cdot\rangle_{-}$ is the inner product of two covariant vectors, and $\langle\cdot,\cdot,\cdot\rangle_{--}$
the matrix product of a covariant matrix with two covariant vectors (see Table~\ref{tab:System_notation}).
Taking the ratio of PI representations (\ref{eq:C_ff}) and (\ref{eq:C_dd(PI)})
of $C_{\text{ff}}$ and $C_{\text{dd}}$ immediately yields the estimator for
the ratio $C_{\text{ff}}/C_{\text{dd}}$:%
\begin{equation}
\left(  \frac{C_{\mathrm{ff}}}{C_{\mathrm{dd}}}\right)_{P,\mathrm{est}%
}=f_{\mathrm{v}}.
\end{equation}
The thermodynamic estimator takes the form\cite{Yamamoto_Miller:2004}
\begin{equation}
f_{\mathrm{v,th}}=-\left(  \frac{P}{2\hbar\beta}\right)  ^{2}\prod
_{\gamma=a,b}\left\{  \left\langle \nabla\xi_{\gamma}(\mathbf{r}_{\gamma
}),(\mathbf{r}_{\gamma}^{(+1)}-\mathbf{r}_{\gamma}^{(-1)})\right\rangle
_{0}%
\vphantom{\left[\nabla\xi_{\gamma}(\mathbf{r}_{\gamma})\cdot(\mathbf{r}_{\gamma}^{(+1)}-\mathbf{r}_{\gamma}^{(-1)})\right]}/||\nabla
\xi_{\gamma}(\mathbf{r}_{\gamma})||_{-}\right\}  \text{,} \label{eq:f_th}%
\end{equation}
where $ \langle\cdot,\cdot\rangle_{0}$ is the inner product of
a covariant and contravariant vectors (see Table \ref{tab:System_notation}),
and we defined
\begin{equation}
\mathbf{r}_{\gamma}^{(+1)}:=%
\begin{cases}
\mathbf{r}^{(P/2+1)}, & \gamma=a,\\
\mathbf{r}^{(1)}, & \gamma=b,
\end{cases}
\end{equation}%
\begin{equation}
\mathbf{r}_{\gamma}^{(-1)}:=%
\begin{cases}
\mathbf{r}^{(P/2-1)}, & \gamma=a,\\
\mathbf{r}^{(P-1)}, & \gamma=b.
\end{cases}
\end{equation}

\subsection{Thermodynamic integration with respect to mass
\label{subsec:Thermodynamic_integration}}

The last ingredient needed for evaluating the QI\ rate constant
\eqref{eq:QI_basic} is the ratio $C_{\text{dd}}/Q_{r}$, which, unfortunately,
cannot be calculated by the standard Metropolis algorithm. However, in the case of
KIE's, one can circumvent this problem by employing the so-called
thermodynamic integration with respect to mass,\cite{Vanicek_Aoiz:2005} which is
easy to understand from the explicit QI expression for the KIE,
\begin{equation}
\mathrm{KIE}_{\text{QI}}=\frac{k^{(A)}_{\mathrm{QI}}}{k^{(B)}_{\mathrm{QI}}}=\frac{Q_{r}^{(B)}}{Q_{r}^{(A)}}
\frac{C_{\text{dd}}^{(A)}}{C_{\text{dd}}^{(B)}}\frac{\Delta H^{(B)}}{\Delta 
H^{(A)}}\frac{C_{\text{ff}}^{(A)}/C_{\text{dd}}^{(A)}}{C_{\text{ff}}
^{(B)}/C_{\text{dd}}^{(B)}}, \label{eq:KIE_QI}%
\end{equation}
where $A$ and $B$ are different isotopologs of otherwise the same system. The
basic idea of the thermodynamic integration with respect to mass
consists in computing ratios $C_{\text{dd}}^{(A)}/C_{\text{dd}%
}^{(B)}$ and $Q_{r}^{(A)}/Q_{r}^{(B)}$ by considering a continuous
 transformation\cite{Vanicek_Aoiz:2005, Zimmermann_Vanicek:2009, Perez_Lilienfeld:2011, Ceriotti_Markland:2013} from $A$ to $B$ using a
dimensionless parameter $\lambda\in\lbrack0,1]$ controlling atomic masses of
the intermediate systems as
\begin{equation}
m_{i}(\lambda)=(1-\lambda)m_{i}^{(A)}+\lambda m_{i}^{(B)}\text{.}%
\end{equation}
Ratios $C_{\text{dd}}^{(A)}/C_{\text{dd}}^{(B)}$ and $Q_{r}^{(A)}/Q_{r}^{(B)}$
are rewritten in terms of their logarithmic derivatives, which are normalized
quantities and, therefore, \emph{can} be calculated with the Metropolis
algorithm:
\begin{equation}
\frac{Q_{r}^{(B)}}{Q_{r}^{(A)}}=\mathrm{exp}\left(  \int_{0}^{1}%
\frac{d\mathrm{ln}Q_{r}^{(\lambda)}}{\mathit{d}\lambda}\mathit{d}%
\lambda\right)  \text{,} \label{eq:Q_r_ratio}%
\end{equation}%
\begin{equation}
\frac{C_{\text{dd}}^{(B)}}{C_{\text{dd}}^{(A)}}=\mathrm{exp}\left(  \int%
_{0}^{1}\frac{d\mathrm{ln}C_{\text{dd}}^{(\lambda)}}{\mathit{d}\lambda
}\mathit{d}\lambda\right)  \text{.} \label{eq:C_dd_ratio}%
\end{equation}
The integrals in the exponent can be evaluated numerically
with Simpson's rule or other standard methods. However, several approaches
have been proposed
for decreasing the exponentiated integration error
in the ratio $ Q_{r}^{(A)}/Q_{r}^{(B)}$, which can further accelerate
the calculation by lowering the required number of integration points: These include 
rescaling of mass (which in the simplest variant involves linearly interpolating $ m^{-1/2}_{i}$ instead of $ m_{i}$),\cite{Ceriotti_Markland:2013, Marsalek_Tuckerman:2014}
or introducing higher-order derivatives of $ Q_{r}$ with respect to $ \lambda$;\cite{Marsalek_Tuckerman:2014}
if the ratio is close to unity it is also possible to eliminate
the integration error altogether by using direct estimators for $ Q_{r}^{(A)}/Q_{r}^{(B)}$.\cite{Cheng_Ceriotti:2014}

In the case of $d\mathrm{ln}C_{\text{dd}}/\mathit{d}\lambda$ needed in the QI,
one needs to keep track of the possible change of
$\xi_{\gamma}$ during the course of the integration,%
\begin{equation}
\frac{d\mathrm{ln}C_{\text{dd}}^{(\lambda)}}{\mathit{d}\lambda}=\frac
{\partial\mathrm{ln}C_{\text{dd}}^{(\lambda)}}{\partial\lambda}+\sum
_{\gamma=a,b}\sum_{k}\frac{\mathit{d}\eta_{k}^{(\gamma)}}{\mathit{d}\lambda
}\frac{\partial\mathrm{ln}C_{\text{dd}}}{\partial\eta_{k}^{(\gamma)}}\text{.}
\label{eq:total_der_Cdd}%
\end{equation}
In Ref. \onlinecite{Vanicek_Aoiz:2005} the authors proposed to choose
$\{\eta_{k}^{(\gamma)}(\lambda)\}$ that satisfy Eq. \eqref{eq:stationarity} at
each $\lambda$ integration step, making the second term in
Eq.~(\ref{eq:total_der_Cdd}) exactly zero and leaving only $\partial
\mathrm{ln}C_{\text{dd}}/\partial\lambda$ to be considered. Here we take an
alternative and more numerically stable approach: By introducing new accurate
estimators for $\partial\mathrm{ln}C_{\text{dd}}/\partial\eta_{k}^{(\gamma)},$
we can avoid having to find the optimal values of $\eta_{k}^{(\gamma)}%
(\lambda)$ for all $\lambda$. Instead, we only find optimal $\eta_{k}%
^{(\gamma)}(\lambda)$ at the boundary points $\lambda\in\{0,1\}$, obtain
other, not necessarily optimal, $\eta_{k}^{(\gamma)}(\lambda)$ by linear
interpolation, and evaluate both terms of Eq. (\ref{eq:total_der_Cdd}) for
each $\lambda$.

The estimators for $\partial\mathrm{ln}Q_{r}/\partial\lambda$ and
$\partial\mathrm{ln}C_{\text{dd}}/\partial\lambda$ are
\begin{align}
\left(  \frac{1}{\beta} \frac{\partial\mathrm{ln}Q_{r}^{(\lambda)}}%
{\partial\lambda}\right)  _{P,\mathrm{est}}  &  =F_{r}=\frac{1}{\beta}%
\sum_{i=1}^{N}\frac{dm_{i}}{d\lambda}\left(  \frac{d\mathrm{ln}C}{dm_{i}%
}-\beta\frac{d\Phi}{dm_{i}}\right)  ,\label{eq:F}\\
\left(  \frac{1}{\beta} \frac{\partial\mathrm{ln}C_{\text{dd}}^{(\lambda)}%
}{\partial\lambda}\right)  _{P,\mathrm{est}}  &  =F^{\ddagger}=F_{r}%
+F_{\mathrm{ds}}, \label{eq:F_ts}%
\end{align}
where $F_{\mathrm{ds}}$ is the contribution that comes from differentiating
the mass-dependent normalization factor in Eq.~\eqref{eq:normalised_delta}:
\begin{equation}
F_{\mathrm{ds}}=-\frac{1}{\beta}\sum_{i=1}^{N}\frac{\mathit{d}m_{i}%
}{\mathit{d}\lambda}\sum_{\gamma=a,b}\frac{|\nabla_{i}\xi_{\gamma}%
(\mathbf{r}_{\gamma})|^{2}}{2m_{i}^{2}||\nabla\xi_{\gamma}(\mathbf{r}_{\gamma
})||_{-}^{2}}.
\end{equation}
Here $\nabla_{i}$ is the gradient with respect to coordinates of particle $i$ (see Table~\ref{tab:System_notation}).

Direct evaluation of Eqs.~\eqref{eq:F} and \eqref{eq:F_ts} yields the
thermodynamic estimators\cite{Vanicek_Aoiz:2005}%
\begin{align}
F_{r,\mathrm{th}}  &  =\frac{1}{\beta}\sum_{i=1}^{N}\frac{\mathit{d}m_{i}%
}{\mathit{d}\lambda}\left[  \frac{DP}{2m_{i}}-\frac{P}{2\hbar^{2}\beta}%
\sum_{s=1}^{P}(\mathrm{r}_{i}^{(s)}-\mathrm{r}_{i}^{(s-1)})^{2}\right]
+F_{r,\mathrm{grad}}\text{,}\label{eq:A_th}\\
F_{\mathrm{th}}^{\ddagger}  &  =F_{r,\mathrm{th}}+F_{\mathrm{ds}}\text{.}
\label{eq:A_ts_th}%
\end{align}
Derivation of the estimator for $\partial\mathrm{ln}C_{\text{dd}}/\partial
\eta_{k}^{(\gamma)}$ involves a rather tedious algebra and is therefore
presented in Appendix B; the result is
\begin{equation}
\left(  \frac{\mathrm{\partial\mathrm{ln}}C_{\mathrm{dd}}}{\partial\eta
_{k}^{(\gamma)}}\right)  _{P,\mathrm{est}}=B^{k(\gamma)}=\frac{\partial
\xi_{\gamma}(\mathbf{r}_{\gamma})}{\partial\eta_{k}^{(\gamma)}}\left\{
\beta\left\langle \nabla\xi_{\gamma}(\mathbf{r}_{\gamma}),\nabla^{(\gamma
)}\Phi(\{\mathbf{r}^{(s)}\})\right\rangle _{-}-B_{\mathrm{ds}}^{k(\gamma
)}\right\}  /||\nabla\xi_{\gamma}(\mathbf{r}_{\gamma})||_{-}^{2},
\label{eq:B_general}%
\end{equation}
where $\nabla^{(\gamma)}$ is the gradient with respect to $\mathbf{r}_{\gamma
}$ and
\begin{equation}
B_{\mathrm{ds}}^{k(\gamma)}=\langle\nabla,\nabla\xi\rangle_{-}-\frac
{1}{||\nabla\xi||_{-}^{2}}\left\langle \nabla\xi,\frac{\partial^{2}\xi
}{(\partial\mathbf{r}_{\gamma})^{2}},\nabla\xi\right\rangle _{--}%
\end{equation}
is the term associated with the change of configuration space volume satisfying the
constraint. Obtaining the thermodynamic estimator for $B^{k(\gamma)}$ is
straightforward and yields
\begin{equation}%
\begin{split}
B_{\mathrm{th}}^{k(\gamma)}=  &  \frac{\partial\xi_{\gamma}(\mathbf{r}%
_{\gamma})} {\partial\eta_{k}^{(\gamma)}}\left\{  \frac{P}{\hbar^{2}\beta
}\left\langle \nabla\xi_{\gamma}(\mathbf{r}_{\gamma}),(2\mathbf{r}_{\gamma
}-\mathbf{r}_{\gamma}^{(-1)}-\mathbf{r}_{\gamma}^{(+1)})\right\rangle
_{0}+w_{\gamma}\left\langle \nabla\xi_{\gamma}(\mathbf{r}_{\gamma}),\nabla
V_{\mathrm{eff}}^{(s)}(\mathbf{r}_{\gamma})\right\rangle _{-}-B_{\mathrm{ds}%
}^{k(\gamma)}\right\} \\
&  /||\nabla\xi_{\gamma}(\mathbf{r}_{\gamma})||_{-}^{2}.
\end{split}
\end{equation}

\subsection{Virial estimators\label{subsec:Centroid-virial-estimators}}

So far we have only considered thermodynamic estimators, which are obtained
via direct differentiation of the Boltzmann operator matrix elements. However,
an estimator for a given quantity is not unique; it is often possible to
obtain an estimator with smaller statistical error. Among such estimators are
the so-called virial and centroid virial estimators,\cite{Herman_Berne:1982,
Parrinello_Rahman:1984} which are motivated by the virial theorem of classical
mechanics and which can be derived\cite{Predescu_Doll:2002, Yamamoto:2005}
most simply by applying a coordinate transformation before the differentiation.

Two of the five virial estimators used in this work, namely the estimators for
$\partial\mathrm{ln}Q_{r}/\partial\lambda$ and $\Delta H^{2}$ had been
proposed previously;\cite{Vanicek_Miller:2007,Yang_Miller:2006} the former,
however, had not been used in combination with the SC
factorization. To derive the centroid virial estimator for $\partial
\mathrm{ln}Q_{r}/\partial\lambda$, let us choose an arbitrary bead $u$ and
rewrite $Q_{r}$ in terms of the coordinates
\begin{equation}
\mathrm{x}_{i}^{(s)}=\mathrm{r}_{i}^{(u)}+\sqrt{m_{i}/m_{i}^{\prime}%
}(\mathrm{r}_{i}^{(s)}-\mathrm{r}_{i}^{(u)}),
\end{equation}
where $\{m_{i}^{\prime}\}$ are a set of parameters with dimensionality of
mass. One then substitutes the new $C$ and $\Phi$ resulting from the
transformation of coordinates into Eq. \eqref{eq:F}, and, finally, sets
$\{m_{i}^{\prime}\}=\{m_{i}\}$ and transforms back to initial coordinates.
This procedure yields an improved virial estimator,
\begin{equation}
F_{r,\mathrm{cv}}^{(u)}=\sum_{i=1}^{N}\frac{1}{2m_{i}}\frac{\mathit{d}m_{i}%
}{\mathit{d}\lambda}\left\{  \frac{D}{\beta}+\frac{1}{P}\sum_{s=1}^{P}%
w_{s}\left[  (\mathrm{r}_{i}^{(s)}-\mathrm{r}_{i}^{(u)})\cdot\nabla
_{i}V_{\mathrm{eff}}^{(s)}(\mathbf{r}^{(s)})\right]
\vphantom{\sum_{s=1}^{P}}\right\}  +F_{r,\mathrm{grad}},
\end{equation}
which, however, depends on an arbitrary choice of bead $u$. After taking the
arithmetic average of all $P$ estimators corresponding to $ P$ different choices of
$u\in\{1,...,P\}$ the centroid virial estimator is obtained,
\begin{equation}
F_{r,\mathrm{cv}}=\sum_{i=1}^{N}\frac{1}{2m_{i}}\frac{\mathit{d}m_{i}%
}{\mathit{d}\lambda}\left\{  \frac{D}{\beta}+\frac{1}{P}\sum_{s=1}^{P}%
w_{s}\left[  (\mathrm{r}_{i}^{(s)}-\mathrm{r}_{i}^{(C)})\cdot\nabla
_{i}V_{\mathrm{eff}}^{(s)}(\mathbf{r}^{(s)})\right]  \right\}  +F_{r,\mathrm{grad}},
\end{equation}
where%
\begin{equation}
\mathbf{r}^{(C)}=\frac{1}{P}\sum_{s=1}^{P}\mathbf{r}^{(s)}%
\end{equation}
is the centroid coordinate. From now on we will refer to this estimator as
``virial''; originally the name ``centroid virial'' was introduced to
distinguish the estimator from the simple virial estimator derived in Ref.
\onlinecite{Vanicek_Miller:2007}, which was not considered in this work since
its statistical error is larger then the error of its centroid counterpart.

For $\Delta H^{2}$, one starts\cite{Yang_Miller:2006} by rewriting Eq.
(\ref{eq:C_dd_zeta}) using the coordinates
\begin{equation}
\mathbf{x}^{(s)}=%
\begin{cases}
\check{\mathbf{r}}^{(s)}+\sqrt{\frac{\beta}{\beta^{+}}}(\mathbf{r}%
^{(s)}-\check{\mathbf{r}}^{(s)}), & 0<s<P/2,\\
\check{\mathbf{r}}^{(s)}+\sqrt{\frac{\beta}{\beta^{-}}}(\mathbf{r}%
^{(s)}-\check{\mathbf{r}}^{(s)}), & P/2<s<P,\\
\mathbf{r}^{(s)}, & s=0,P/2,P,
\end{cases}
\label{eq:resc_for_dH_v}%
\end{equation}
where $\check{\mathbf{r}}^{(s)}$ is the reference point given by%
\begin{equation}
\check{\mathbf{r}}^{(s)}=\check{\mathbf{r}}^{(P-s)}=\mathbf{r}^{(P)}%
+(\mathbf{r}^{(P/2)}-\mathbf{r}^{(P)})\frac{s}{P/2}~~~(0<s<P/2).
\end{equation}
The kinetic parts of $\tilde{\Phi}^{\pm}$ are rewritten in the new
coordinates; e.g., for $\tilde{\Phi}^{+}$, one uses the relation
\begin{equation}
\frac{1}{\beta^{+}}\sum_{s=1}^{P/2}||\mathbf{r}^{(s)}-\mathbf{r}^{(s-1)}%
||_{+}^{2}=\frac{1}{\beta}\sum_{s=1}^{P/2}||\mathbf{x}^{(s)}-\mathbf{x}%
^{(s-1)}||_{+}^{2}+\left(  \frac{1}{\beta^{+}}-\frac{1}{\beta}\right)
\frac{||\mathbf{x}^{(P/2)}-\mathbf{x}^{(P)}||_{+}^{2}}{P/2}.
\end{equation}
By substituting transformed $\tilde{\Phi}^{\pm}$ and $\overline{C}$ into Eqs.
\eqref{eq:G_factor} and \eqref{eq:F_factor}, one obtains the desired $G$ and
$F$ terms of the virial estimator:
\begin{equation}%
\begin{split}
G_{\mathrm{v}}=  &  \vphantom{*}\frac{4ND}{\beta^{2}}-\frac{16}{\hbar^{2}%
\beta^{3}}||\mathbf{r}^{(P/2)}-\mathbf{r}^{(P)}||_{+}^{2}\\
&  -\frac{1}{\beta P}\sum_{s=1}^{P}w_{s}\left[  3\left\langle (\mathbf{r}%
^{(s)}-\check{\mathbf{r}}^{(s)}),\nabla V_{\mathrm{eff}}^{(s)}(\mathbf{r}%
^{(s)})\right\rangle _{0}%
\vphantom{\left \langle\frac{\partial^2 V_{\mathrm{eff}}^{(s)}}{(\partial \mathbf{r}^{s})^{2}}\right\rangle_{0}}\right.
\\
&  \left.  +\left\langle (\mathbf{r}^{(s)}-\check{\mathbf{r}}^{(s)}%
),\frac{\partial^{2}V_{\mathrm{eff}}^{(s)}(\mathbf{r}^{(s)})}{(\partial
\mathbf{r}^{(s)})^{2}},(\mathbf{r}^{(s)}-\check{\mathbf{r}}^{(s)}%
)\right\rangle _{00}\right] \\
&  +G_{\mathrm{v,grad}},
\end{split}
\end{equation}%
\[
F_{\mathrm{v}}=\frac{2}{P}\left(  \sum_{s=1}^{P/2-1}-\sum_{s=P/2+1}%
^{P-1}\right)  w_{s}\left[  V_{\mathrm{eff}}^{(s)}(\mathbf{r}^{(s)})+\frac
{1}{2}\left\langle (\mathbf{r}^{(s)}-\check{\mathbf{r}}^{(s)}),\nabla
V_{\mathrm{eff}}^{(s)}(\mathbf{r}^{(s)})\right\rangle _{0}\right]
+F_{\mathrm{grad}},
\]
where $\langle\cdot,\cdot,\cdot\rangle_{00}$ is
the matrix product of a covariant matrix with two contravariant vectors (see Table~\ref{tab:System_notation}).

Now let us turn to the derivation of the new estimators promised in the
Introduction. In particular, we propose new virial estimators for
$\partial\mathrm{ln}C_{\mathrm{dd}}/\partial\lambda$, $C_{\mathrm{ff}%
}/C_{\mathrm{dd}}$, and $\partial\mathrm{ln}C_{\text{dd}}/\partial\eta
_{k}^{(\gamma)}$. For $\partial\mathrm{ln}C_{\text{dd}}/\partial\lambda$ we
use a coordinate rescaling
\begin{equation}
\mathrm{x}_{i}^{(s)}=\check{\mathrm{r}}_{i}^{(s)}+\sqrt{m_{i}/m_{i}^{\prime}%
}(\mathrm{r}_{i}^{(s)}-\check{\mathrm{r}}_{i}^{(s)}),
\end{equation}
which is similar to Eq. \eqref{eq:resc_for_dH_v} and yields the virial
estimator%
\begin{equation}%
\begin{split}
F_{\mathrm{cv}}^{\ddagger}=  &  \sum_{i=1}^{N}\frac{\mathit{d}m_{i}%
}{\mathit{d}\lambda}\left\{  \frac{D}{\beta m_{i}}-\frac{2}{(\beta\hbar)^{2}%
}(\mathrm{r}_{i}^{(P/2)}-\mathrm{r}_{i}^{(P)})^{2}+\frac{1}{2Pm_{i}}\sum
_{s=1}^{P}w_{s}\left[  (\mathrm{r}_{i}^{(s)} -\check{\mathrm{r}}_{i}%
^{(s)})\cdot\nabla_{i}V_{\mathrm{eff}}^{(s)}(\mathbf{r}^{(s)})\right]
\right\} \\
&  +F_{\mathrm{ds}}+F_{r,\mathrm{grad}}.
\end{split}
\end{equation}

For $C_{\text{ff}}/C_{\text{dd}}$, we introduce new coordinates
\begin{equation}
\mathbf{x}^{(s)}=\mathbf{r}^{(s)}-\check{\mathbf{r}}^{(s)}%
\end{equation}
and employ the identity%
\begin{equation}
\sum_{s=1}^{P/2}||\mathbf{r}^{(s)}-\mathbf{r}^{(s-1)}||_{+}^{2}=\sum
_{s=1}^{P/2}||\mathbf{x}^{(s)}-\mathbf{x}^{(s-1)}||_{+}^{2}+\frac
{||\mathbf{r}^{(P/2)}-\mathbf{r}^{(P)}||_{+}^{2}}{P/2}.
\end{equation}
Rewriting $\tilde{\Phi}^{\pm}$ in terms of $\{\mathbf{x}^{(1)},$ $...,$
$\mathbf{x}^{(P/2-1)},$ $\mathbf{r}^{(P/2)},$ $\mathbf{x}^{(P/2+1)},$ $...,$
$\mathbf{x}^{(P-1)},$ $\mathbf{r}^{(P)}\}$ and inserting them into Eq.
\eqref{eq:f_general} leads to the virial estimator
\begin{align}
f_{\mathrm{v,v}}  &  =(\beta^{2}v_{a}v_{b}-g_{ab})/\{\prod_{\gamma
=a,b}||\nabla\xi_{\gamma}(\mathbf{r}_{\gamma})||_{-}\},\label{eq:f_v}\\
v_{\gamma}  &  =\frac{\hbar}{P^{2}}\left(  \sum_{s=1}^{P/2-1}-\sum
_{s=P/2+1}^{P-1}\right)  e^{(s)}_{\gamma}w_{s}\left\langle \nabla\xi_{\gamma
}(\mathbf{r}_{\gamma}),\nabla V_{\mathrm{eff}}^{(s)}(\mathbf{r}^{(s)}%
)\right\rangle _{-},\\
g_{ab}  &  =\frac{\hbar^{2}\beta}{P^{3}}\sum_{s=1}^{P}e^{(s)}_{a}e^{(s)}%
_{b}w_{s}\left\langle \nabla\xi_{a}(\mathbf{r}^{(P/2)}),\frac{\partial
^{2}V_{\mathrm{eff}}^{(s)}(\mathbf{r}^{(s)})}{(\partial\mathbf{r}^{(s)})^{2}%
},\nabla\xi_{b}(\mathbf{r}^{(P)})\right\rangle _{--}-\frac{\left\langle
\nabla\xi_{a}(\mathbf{r}^{(P/2)}),\nabla\xi_{b}(\mathbf{r}^{(P)})\right\rangle
_{-}}{\beta},
\end{align}
where we introduced coefficients
\begin{equation}
e^{(s)}_{\gamma}=%
\begin{cases}
\mathrm{min}(s,P-s), & \gamma=a,\\
|s-P/2|, & \gamma=b.
\end{cases}
\end{equation}

Using the same rescaling as for $f_{\mathrm{v,v}}$, we can also derive the
virial estimator for $\partial\mathrm{ln}C_{\text{dd}}/\partial\eta
_{k}^{(\gamma)}$,
\begin{equation}%
\begin{split}
B_{\mathrm{v}}^{k(\gamma)}=  &  \frac{\partial\xi_{\gamma}}{\partial\eta
_{k}^{(\gamma)}}\left\{  \frac{4}{\hbar^{2}\beta}\left\langle \nabla
\xi_{\gamma}(\mathbf{r}_{\gamma}),(\mathbf{r}_{\gamma}-\mathbf{r}_{\gamma
}^{(P/2)})\right\rangle _{0}\vphantom{\sum_{s=1}^{P}}\right. \\
&  \left.  +\frac{2\beta}{P^{2}}\sum_{s=1}^{P}e^{(s)}_{\gamma}\left\langle
\nabla\xi_{\gamma}(\mathbf{r}_{\gamma}),\nabla V_{\mathrm{eff}}^{(s)}%
(\mathbf{r}^{(s)})\right\rangle _{-}%
\vphantom{\left\{\sum_{s=1}^{P}\right\}}-B_{\mathrm{ds}}^{k(\gamma)}\right\}
\\
&  /||\nabla\xi_{\gamma}||_{-}^{2},
\end{split}
\end{equation}
where $\mathbf{r}_{\gamma}^{(P/2)}$ stands for $\mathbf{\mathbf{r}}^{(P)}$ if
$\gamma=a$ and for $\mathbf{\mathbf{r}}^{(P/2)}$ if $\gamma=b$.

We would like to comment on the cost of using the estimators described in this
subsection. While thermodynamic estimators require little numerical effort,
their virial counterparts depend on the gradient and Hessian of the
effective potential. (Note that although $B_{\mathrm{th}}^{k(\gamma)}$ also
depends on the force, it depends only on the force acting on a single bead,
and hence its cost is negligible for large $P$.) It should be emphasized,
however, that gradient- and Hessian-dependent parts of virial estimators can
be calculated by finite difference, without the need to evaluate the gradient
or Hessian explicitly. For example,
$ \langle\mathbf{w},\nabla V\rangle_{0}$ and $ \langle\mathbf{w},\partial^{2} V/\partial
\mathbf{r}^{2}, \mathbf{w}\rangle_{00}$
are first and second derivatives of $ V$ in the direction of
$ \mathbf{w}$, and therefore can be evaluated by finite difference using just one and two
additional evaluations of $ V$,
respectively. As a result, the effective cost is only
one extra potential
evaluation per bead for $F_{r,\mathrm{cv}}$, one per unconstrained bead for
$F_{\mathrm{cv}}^{\ddagger}$, two per unconstrained bead for $(G_{\mathrm{v}%
}+F_{\mathrm{v}}^{2})/2$, and three for $f_{\mathrm{v,v}}$. Calculating
$B_{\mathrm{th}}^{k(\gamma)}$ will require exactly one potential evaluation
and calculating $B_{\mathrm{v}}^{k(\gamma)}$ will require $P-1$ evaluations
unless it is computed at the same time as $f_{\mathrm{v,v}}$ (in this case it
would require just one extra potential evaluation, other numerical ingredients
being shared with $f_{\mathrm{v,v}}$).

It should be emphasized that it is not necessary to evaluate these estimators
after each MC step due to finite correlation lengths inherent to MC
simulations. This realization frequently allows one to make the additional
cost of evaluating even the more expensive estimators small compared with the
cost of the random walk itself.

Finally, we would like to point out that, while authors of Refs.
\onlinecite{Vanicek_Miller:2007} and \onlinecite{Yamamoto:2005} used finite
differences with respect to mass and $\beta$, respectively,
to calculate virial estimators of interest, we found this approach less convenient since it
requires introducing two parameters (finite difference steps) that must be
adjusted for each new isotopolog and for each temperature. We therefore only
used finite differences with respect to coordinates in the system's
configuration space, with a single finite difference step which is the same
for all isotopologs and all temperatures.

\section{Applications\label{sec:Application}}

In summary, to compute the KIE on a reaction one must:

1. Estimate the Trotter number $P$ that is sufficient to adequately describe
the system. For this purpose we made several preliminary calculations to
estimate the $P$ necessary for the lowest and highest temperature; for other
temperatures we used the empirical rule that $1/P$ stays approximately linear
with respect to $T$.

2. Choose the two DS's. We chose $\xi_{\gamma}(\mathbf{r})$ of
the form
\begin{equation}
\xi_{\gamma}(\mathbf{r})=\xi(\mathbf{r})-\eta_{\gamma}^{\ddagger}\text{.}
\label{eq:xi_used_here}%
\end{equation}
For reactions where atom \textit{X} breaks its bond with atom \textit{Y} and
forms a bond with atom \textit{Z}, we used as a reaction coordinate the
difference of the \textquotedblleft bond\textquotedblright\ lengths, i.e.,
\begin{equation}
\xi(\mathbf{r})=R_{XY}-R_{XZ}\text{,} \label{eq:reaction_coordinate}%
\end{equation}
where $R_{XY}$ is the distance between \textit{X} and \textit{Y}. Optimal
values of $\eta_{\gamma}^{\ddagger}$ were found by running test simulations to
find the sign of $\partial\mathrm{ln}C_{\text{dd}}/\partial\eta_{\gamma
}^{\ddagger}$ at different values of $(\eta_{a}^{\ddagger},\eta_{b}^{\ddagger
})$.

3. Run simulations at different values of $\lambda$ in order to obtain the
corresponding logarithmic derivatives of $Q_{r}$ and $C_{\text{dd}}$, as well
as $C_{\text{ff}}(0)/C_{\text{dd}}(0)$ and $\Delta H$ for $\lambda=0$ and
$\lambda=1$, then evaluate Eqs. \eqref{eq:Q_r_ratio} and \eqref{eq:C_dd_ratio}
using, e.g., Simpson's rule. For many systems $d\mathrm{ln}C_{\text{dd}%
}/\mathit{d}\lambda$ and $\partial\mathrm{ln}Q_{r}/\partial\lambda$ are quite
smooth functions and nine intermediate points were sufficient to accurately
evaluate the thermodynamic integrals (i.e., the discretization error of the
$\lambda$ integral was smaller than the already small statistical error).
After this, evaluating the KIE using Eq. \eqref{eq:KIE_QI} is straightforward.

For each value of $\lambda$ one has to run two MC
simulations in $\{\mathbf{r}^{(s)}\}$: a \textquotedblleft constrained
simulation\textquotedblright\ with two slices constrained to their respective
DS and a standard (\textquotedblleft
unconstrained\textquotedblright) simulation. Since treating exact constraints
is not straightforward in MC methods, we approximated the delta
constraint with a \textquotedblleft smeared\textquotedblright\ delta function
$\delta_{\text{sm}}$,
\begin{equation}
\delta\lbrack\xi_{\gamma}(\mathbf{r})]\approx\delta_{\text{sm}}[\xi_{\gamma
}(\mathbf{r})]=\frac{1}{\sqrt{2\pi}\sigma}\frac{1}{|\nabla\xi_{\gamma
}(\mathbf{r})|}\exp\left\{  -\frac{1}{2\sigma^{2}}\left[  \frac{\xi_{\gamma
}(\mathbf{r})}{|\nabla\xi_{\gamma}(\mathbf{r})|}\right]  ^{2}\right\}  .
\end{equation}
In contrast with the approximation used in Ref.
\onlinecite{Yamamoto_Miller:2004}, the width $\sigma$ of our Gaussian
$\delta_{m}$ does not depend on temperature or mass. The approximate
constraint converges to the exact delta function as $\sigma\rightarrow0$.
Presence of $\delta_{\text{sm}}[\xi_{\gamma}(\mathbf{r})]$ can be easily
simulated by adding an extra constraining potential to two of the slices. For
MC sampling, we employed the staging
algorithm\cite{Sprik_Chandler:1985,Sprik_Chandler:1985_1,Tuckerman_Klein:1993}
with multislice moves in combination with whole-chain moves. For constrained
simulations, we also made extra single-slice moves of slices $P/2$ and $P$,
since these slices are more rigid than others due to the presence of the
constraining potential.

\subsection{$\mathbf{H+H_{2}}$ rearrangement}

The errors of PI MC calculations come mostly from
two sources: the PI discretization error (due to $P$ being finite)
and the statistical error inherent to MC methods. (As for quantities
evaluated with thermodynamic integration, there is an additional
discretization error of the thermodynamic integral due to taking a finite
number of $\lambda$ steps.) To verify the improvements outlined in
Sec.~\ref{sec:PI_implementation} we studied their influence on the behavior of
the two main types of errors when applied to the model $\mathrm{\cdot
H_{\alpha}+H_{\beta}H_{\gamma}\rightarrow H_{\alpha}H_{\beta}+\cdot H_{\gamma
}}$ rearrangement using the BKMP2 potential energy surface\cite{BKMP2:1996} at
the temperature of $200\,\text{K}$. The behavior of the logarithmic
derivatives was studied on the KIE $\mathrm{\cdot H+H_{2}/\cdot D+D_{2}}$.

\subsubsection{Computational details}

Statistically converged simulations (paralleled over 64 trajectories,
$4\times10^{7}$ MC steps each) were run with different values of the Trotter
number (from $P=8$ to $64$ with step $4$ and from $64$ to $352$ with step
$16$) and different Boltzmann operator factorizations. Virial estimators were
evaluated only after every $25$ MC steps, whereas the thermodynamic -
after every step, because the additional cost was negligible. To estimate
statistical errors of the results we calculated root mean square deviations of
averages over different trajectories. [Having a relatively high number ($64$)
of uncorrelated trajectories, we could thus avoid a more tedious
block-averaging procedure,\cite{Flyvbjerg_Petersen:1989} but we did check in
several cases that the two approaches gave very similar statistical error
estimates.] As for the positions of the DS's, for calculating the
KIE choosing $\eta_{a}^{\ddagger}=\eta_{b}^{\ddagger}=0$ was quite
satisfactory even at $T=200$ K (in this case $C_{\text{dd}}$ is stationary
from symmetry considerations) for analyzing numerical behavior of
$\partial\mathrm{ln}C_{\text{dd}}/\partial\lambda$, $\Delta H^{2}$ and
$C_{\mathrm{ff}}/C_{\mathrm{dd}}$. For $\partial\mathrm{ln}C_{\text{dd}%
}/\partial\eta_{a}^{\ddagger}$, however, we used $\eta_{a}^{\ddagger}=-0.5$
and $\eta_{b}^{\ddagger}=0.5$ in order to make the logarithmic derivative
statistically relevant.

For this particular setup the increase of central processing unit (CPU) time associated with evaluating
all virial estimators at once was about $15\%$ for constrained and $3.5\%$ for
unconstrained simulations. The increase of CPU time associated with the use of
higher-order splittings was negligible for constrained simulations; for
unconstrained simulations it was $2.5\%$ and $5\%$ for SC and TI
splittings, respectively.

\subsubsection{Results}

\begin{figure}
[ptbh]\includegraphics[width=\textwidth]{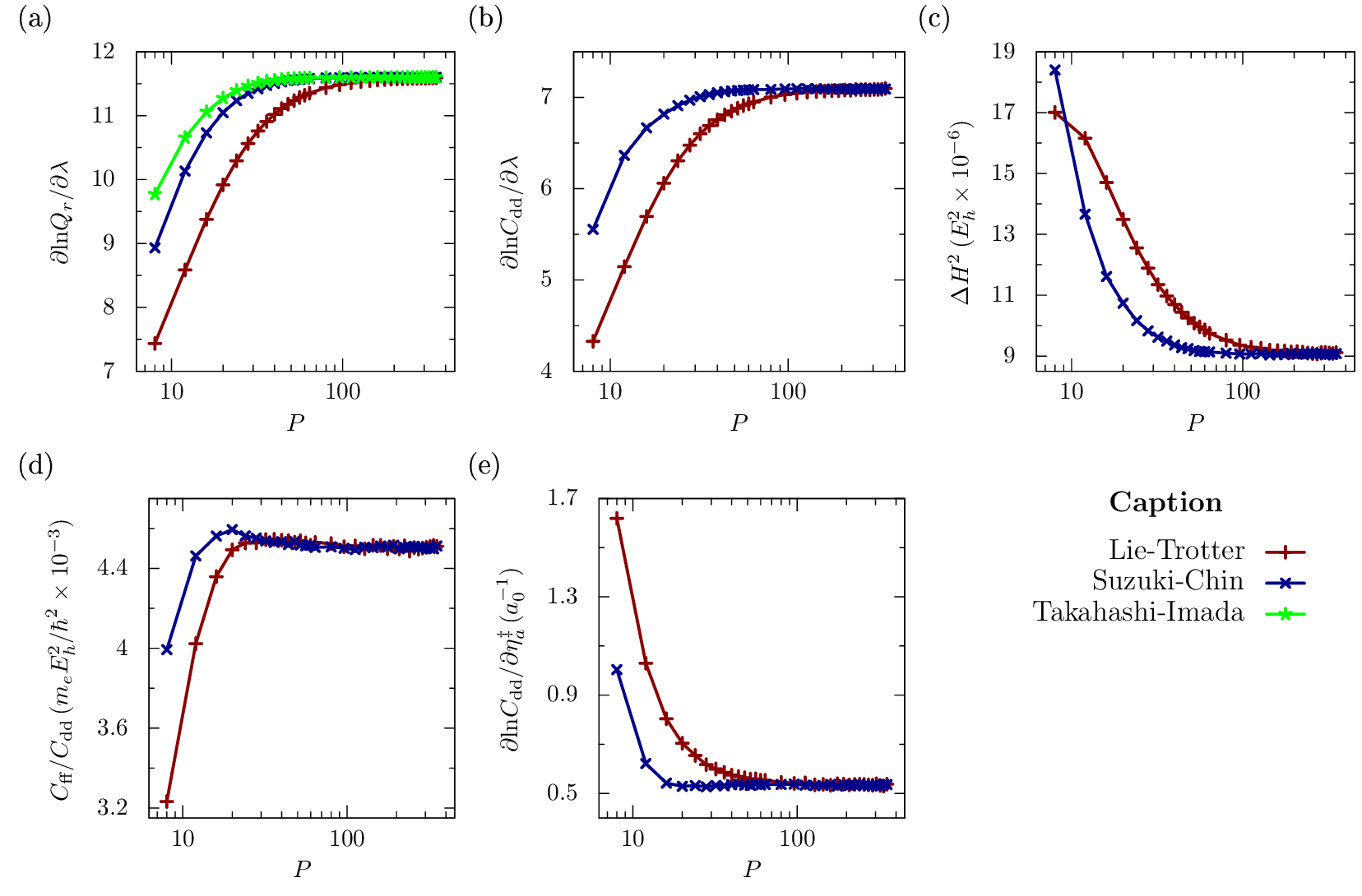}
\caption{Convergence of various quantities required in the QI approximation of the KIE to the
quantum limit as a function of the Trotter number $ P $: (a) $\partial
\mathrm{ln} Q_{r}/\partial\lambda$, (b) $\partial\mathrm{ln}C_{\text{dd}}/\partial\lambda$, (c) $\Delta H^{2}$, (d)
$C_{\mathrm{ff}}/C_{\mathrm{dd}}$, (e)
$\partial\mathrm{ln} C_{\text{dd}}/\partial\eta_{a}^{\ddagger}$.
Results shown were obtained with the virial estimators and correspond to the KIE
$\mathrm{\cdot H+H_{2}/\cdot D+D_{2}}$ at 200 K.\hfill \label{fig:bead_convergence}}
\end{figure}

Convergence of different quantities to their quantum limits as a function of
the Trotter number $P$ is shown in Fig. \ref{fig:bead_convergence}. As
expected, the SC factorization allows to lower the Trotter
number significantly in comparison with the LT factorization. In the
case of $\partial\mathrm{ln}Q_{r}/\partial\lambda$ the SC splitting is
slightly outperformed by the TI factorization, which
has a smaller prefactor of the error term, possibly because the
TI splitting leads to an expression invariant under cyclic bead permutations.

\makeatletter

\makeatother

\begin{figure}
[ptbh]\includegraphics[width=\textwidth]{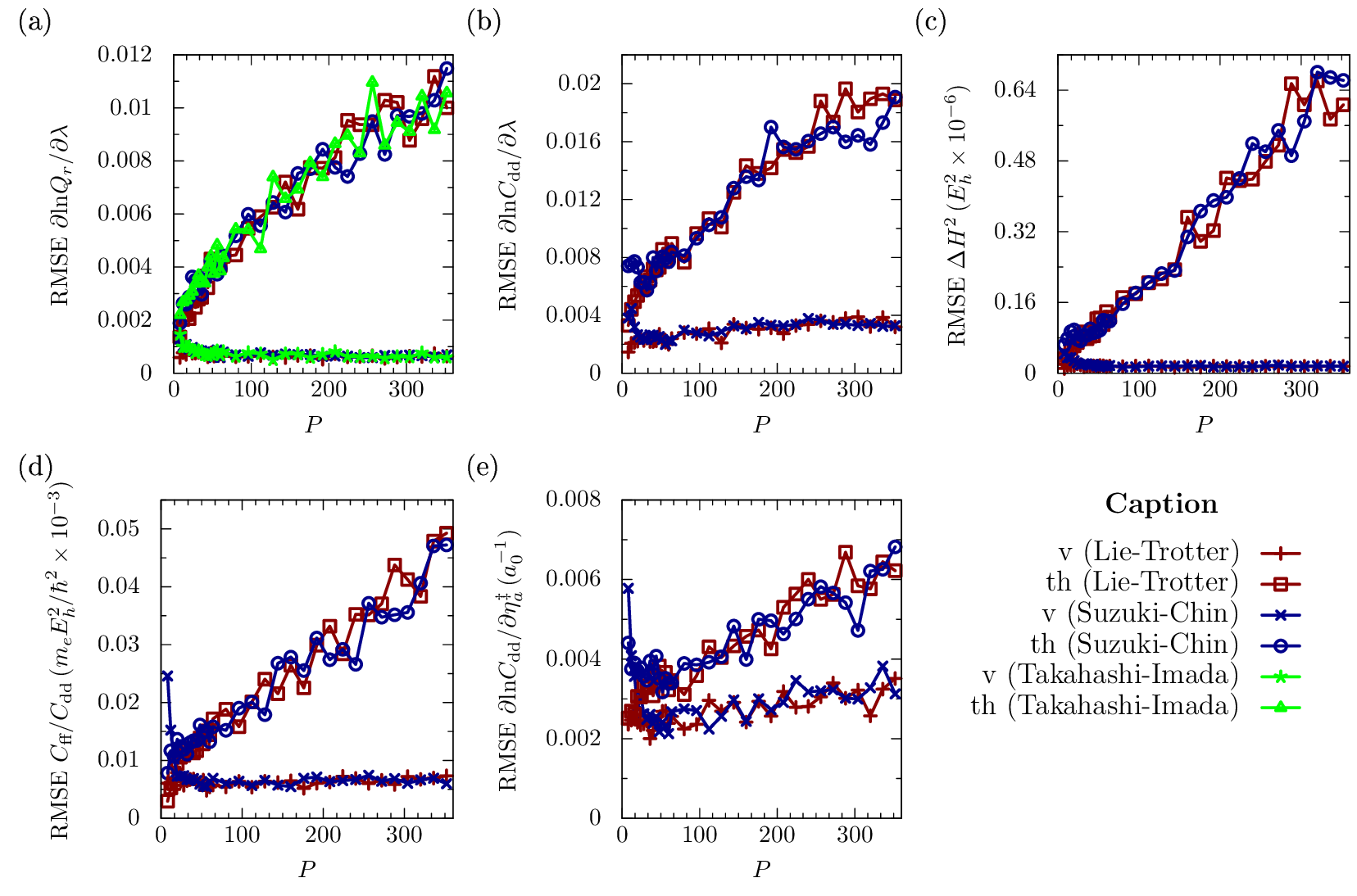}
\caption{Statistical root mean square errors (RMSE) obtained with different
estimators at different values of the Trotter number $P$ for quantities
required in the QI approximation. (a) $\partial\mathrm{ln} Q_{r}/\partial
\lambda$, (b) $\partial\mathrm{ln} C_{\text{dd}}/\partial\lambda$, (c) $\Delta H^{2}$, (d)
$C_{\text{ff}}/C_{\text{dd}}$, (e)
$\partial\mathrm{ln} C_{\text{dd}}/\partial\eta_{a}^{\ddagger}$. Results correspond to the KIE
$\mathrm{\cdot H+H_{2}/\cdot D+D_{2}}$ at 200 K. ``v'' stands for ``virial'', ``th'' - for
``thermodynamic''.\hfill}\label{fig:Statistical_errors}
\end{figure}

Statistical errors of different estimators are presented in Fig.
\ref{fig:Statistical_errors}. Note that they do not depend much on the
factorization used. In contrast, the decrease of statistical errors associated with using
virial estimators is remarkable for all quantities.

\makeatletter

\makeatother

To compare the speedups achieved by different combinations of splittings and estimators we
estimated the relative CPU times needed to converge the quantities
$\Delta H$, $C_{\mathrm{ff}}/C_{\mathrm{dd}}$, $Q_{r}^{(B)}/Q_{r}^{(A)}$, and
$C_{\text{dd}}^{(B)}/C_{\text{dd}}^{(A)}$ to $ 1\%$ discretization and statistical errors.
\footnote{Note that the $1\%$
error for $\Delta H$ translates into a $2\%$ error for $\Delta H^{2}$ and that
$1\%$ relative error for $Q_{r}$ and $C_{\text{dd}}$ ratios translate into
$0.01$ absolute error for $\partial\ln Q_{r}/\partial\lambda$ and $\partial\ln
C_{\mathrm{dd}}/\partial\lambda$. As for $\partial\ln C_{\mathrm{dd}}%
/\partial\eta_{a}^{\ddagger}$, as will be shown later, when we calculate the
KIE $\mathrm{\cdot H+H_{2}/\cdot D+D_{2}}$ at $T=200\,\text{K}$ with a
properly optimized DS, $\partial\ln C_{\mathrm{dd}}/\partial\eta_{a}%
^{\ddagger}$ is integrated over an interval of the length $0.59$ a. u.,
implying that the target error should be $0.01/0.59(\text{a.\thinspace
u.})^{-1}$.} To estimate the speedup associated with calculating the overall KIE itself
with $ 1\%$ statistical and discretization errors
we ran a separate set of simulations with $ \lambda=1$ in addition to those for $ \lambda=0$;
the statistical and discretization errors of the KIE calculated with different combinations of estimators
and factorizations were then approximated with the corresponding errors
obtained if thermodynamic integration of $ Q_{r}$ and $ C_{\mathrm{dd}}$ had been performed using a single step trapezoidal rule
(i.e., based just on the two boundary points $ \lambda=0$ and $ \lambda=1$).

Let us assume the CPU time of a simulation to be approximately
proportional to $P$ and the
number of MC steps. Then for a given combination of factorization and estimator
the cost of achieving the target discretization and statistical
errors is proportional to the product $\tilde{P}\sigma_{\tilde{P}}^{2}$, where $\tilde{P}$
is the value of the Trotter number that yields the
target discretization error and $\sigma_{\tilde{P}}$ is the statistical error
exhibited by the estimator at this value of $P$. These estimates of
CPU\ cost are then corrected by the increase in CPU time associated with using
the fourth-order splittings and virial estimators.
The final results are presented in
Tab. \ref{tab:improvement}, which confirms that the combination of
virial estimators and fourth-order splittings leads to a significant speedup
of the calculation.

One may be surprised that the value of $ P $ necessary to achieve $ 1\% $ convergence of $ C_{\mathrm{ff}}/C_{\mathrm{dd}}$
appears to be roughly independent of the splitting used; this is probably because the discretization errors
of $ C_{\mathrm{dd}}$ and $ C_{\mathrm{ff}}$ cancel to a larger extent for the LT than the
SC splitting. Taking the discretization error to be $ 0.5\%$ rather than $ 1\%$
makes the difference in the required value of $ P$ even
more pronounced: $ P=40$ for the LT and $ P=80$ for the
SC splitting.

Note that even though the values of $ P $ required to converge individual quantities
are quite large (up to $ P=336$ for $ \partial\ln Q_{r}/\partial\lambda$ if LT splitting is used),
the Trotter number $ P $ necessary to converge the final KIE result is significantly lower due to the cancellation of
discretization errors between individual quantities and especially between the two isotopologs.
However, our $ P $ value required for the KIE computed with the LT splitting is still larger than, for instance, those used in
Ref. \onlinecite{Yamamoto_Miller:2004}, where
the authors obtained the final result by extrapolating to the $ P\rightarrow\infty$
limit.\footnote{Since thermodynamic estimators were used in Ref. \onlinecite{Yamamoto_Miller:2004}, reducing the 
discretization error directly using very large $ P $
was not feasible---increasing $ P $ not only decreased discretization error, but also
increased the statistical error. Introducing
virial estimators for each relevant quantity allows avoiding this issue
because it permits improving convergence with respect to $ P$ without
encountering problems with statistical error.}

It is also interesting to relate our results to those of Ref. \onlinecite{Engel_Major:2015},
where the authors compared efficiencies of the LT, TI,
and fourth-order Chin\cite{Chin_Chen:2002, Chin:2004} factorizations for
finding different quantities associated with the RPMD
expression for the reaction rate.
The authors found that for dynamical properties the TI splitting gives little improvement over the standard LT
factorization, which
is consistent with our explanation presented in Subsec. \ref{subsec:Boltzmann_splittings}; both factorizations are
outperformed by the fourth-order Chin factorization, which
is in agreement with the SC outperforming LT splitting in Tab. \ref{tab:improvement}. For equilibrium properties,
the authors found that the efficiencies of the Chin and TI factorizations are similar,
and that both fourth-order factorizations significantly outperform the standard LT splitting,
again in agreement with our results and explanation.

We mentioned earlier that we had calculated virial estimators by finite difference,
making the computational cost of their evaluation independent of dimensionality. To employ fourth-order splittings, however,
one must know the potential gradient for all $ P $ replicas (for the TI splitting)
or at least for $ P/2$ replicas (for the SC splitting if $ \alpha=0$ or $ \alpha=1$).
In general, if evaluating the gradient becomes too expensive compared to the potential energy itself, it may be advantageous
to use the LT instead of the fourth-order splittings.
For example, as shown in Tab. \ref{tab:improvement}, using the fourth-order splittings
decreased the necessary $ P $ approximately four times;
therefore, for this particular system it is reasonable to use the TI factorization if the cost of evaluating
the gradient is smaller than three times the cost of evaluating the potential alone.
For the SC factorization the corresponding factor is around six, since one needs
only $ P/2$ force evaluations. This upper bound for efficiency may be pushed further
 using the reweighting-based techniques;\cite{Jang_Voth:2001, Yamamoto:2005, Perez_Tuckerman:2011}
this approach, however, is known to increase the statistical errors of the final result
in high-dimensional systems.\cite{Ceriotti_Manolopoulos:2012}

\begin{table}
\caption{Estimated speedups of the QI calculations achieved by the use of various combinations of path integral factorizations
and estimators (th = thermodynamic, v = virial). Speedup ``1'' (i.e., no speedup) corresponds to the standard method employing a combination of the
Lie-Trotter factorization and thermodynamic estimators. Results correspond to the KIE
$\mathrm{\cdot H+H_{2}/\cdot D+D_{2}}$ at 200 K.\hfill}\label{tab:improvement}
\begin{ruledtabular}
\begin{tabular}{cdddddddD{,}{}{3.1}dccc}
Factorization &  & \multicolumn{3}{c}{Lie-Trotter (LT)} &  & \multicolumn{3}{c}{Suzuki-Chin (SC)} &  & \multicolumn{3}{c}{Takahashi-Imada (TI)}\tabularnewline
\hline
Estimator &  & \multicolumn{1}{c}{$P$} & \multicolumn{1}{c}{th} & \multicolumn{1}{c}{v} &
& \multicolumn{1}{c}{$ P $} & \multicolumn{1}{c}{th} & \multicolumn{1}{c}{v} &  & \multicolumn{1}{c}{$ P $}
& \multicolumn{1}{c}{th} & \multicolumn{1}{c}{v}\tabularnewline
\hline
$\partial \mathrm{ln} Q_{r}/\partial\lambda$ &  & \multicolumn{1}{>{\centering}p{0.75cm}}{336} & 1 & 220 &  & 96 & 10 & 850, &  & 64 & 44 & 1200\tabularnewline
$\partial \mathrm{ln} C_{\text{dd}}/\partial\lambda$ &  & 240 & 1 & 19 &  & 64 & 13 & 160, &  & \multicolumn{1}{c}{\_}
& \multicolumn{1}{c}{\_} & \multicolumn{1}{c}{\_}\tabularnewline
$\Delta H^{2}$ &  & 128 & 1 & 170 &  & 48 & 14 & 380, &  & \multicolumn{1}{c}{\_} & \multicolumn{1}{c}{\_} &
\multicolumn{1}{c}{\_}\tabularnewline
$C_{\text{ff}}/C_{\text{dd}}$ &  & 36 & 1 & 2.7 &  & 32 & 1.3 & 3,.0 &  & \multicolumn{1}{c}{\_} &
\multicolumn{1}{c}{\_} & \multicolumn{1}{c}{\_}\tabularnewline
$\partial \mathrm{ln} C_{\text{dd}}/\partial\eta_{a}^{\ddagger}$ &  & 80 & 1 & 1.7 &  & 16 & 3.6 & 3,.8 &  & \multicolumn{1}{c}{\_}
& \multicolumn{1}{c}{\_} & \multicolumn{1}{c}{\_}\tabularnewline
KIE &  & 128 & 1 & 34 &  & 40 & 12 & 97,\footnotemark[1] &  & \multicolumn{1}{c}{\_} & \multicolumn{1}{c}{\_}
& \multicolumn{1}{c}{\_}\tabularnewline
\end{tabular}
\footnotetext[1]{All quantities except for $ \partial\ln Q_{r}/\partial\lambda$ are calculated with the SC factorization.
For $ \partial\ln Q_{r}/\partial\lambda$ TI factorization is used.}
\end{ruledtabular}

\end{table}

Lastly, we verified the modified methodology by comparing our result for the
KIE $\mathrm{\cdot H+H_{2}/\cdot D+D_{2}}$ with those of Ref.
\onlinecite{Vanicek_Aoiz:2005}, obtained both with the QI approximation and with an
exact quantum method. For each temperature we calculated $\Delta H$ and
$C_{\text{ff}}/C_{\text{dd}}$ by PIMC simulations with $1.28\times10^{8}$
steps at $\lambda=0$ and $\lambda=1$. Ratios of $Q_{r}$ and $C_{\text{dd}}$
were evaluated by rewriting them as in Eqs. (\ref{eq:Q_r_ratio}) and
(\ref{eq:C_dd_ratio}) respectively and finding the integral over $\lambda$
using Simpson's rule with integration step $\Delta\lambda=0.1$. At
$T=200\ \text{K}$ we also ran calculations with $\Delta\lambda=0.05$ to verify
that the integration error of the final result is lower than the statistical
error. Values of $\partial\mathrm{ln}C_{\text{dd}}/\partial\lambda$ within the
integration interval were obtained by running simulations with $6.4\times
10^{7}$ MC steps (i.e., fewer steps than for the $\lambda$-endpoint
simulations because $\partial\mathrm{ln}C_{\text{dd}}/\partial\lambda$ and
$\partial\mathrm{ln}Q_{r}/\partial\lambda$ tend to converge faster than
$C_{\text{ff}}/C_{\text{dd}}$ and especially than $\Delta H$). These
conditions ensured that the total relative error of the final KIE caused by
statistical noise was below $1\%$. We chose $P$ in such a way that the
relative error due to $P$ being finite was less than the statistical one. At
the lowest temperature $T=200\,\text{K}$ we chose $P=64$, while for
$T=2400\ \text{K}$ $P=12$ turned out to be appropriate; for other temperatures
we estimated the necessary $P$ by interpolation assuming that the $1/P$ is a
linear function of $T$. To verify that the chosen values of $P $ were
sufficient we ran additional simulations at temperatures $200$ K, $1000$ K,
and $2400$ K with $\lambda=0$ and $\lambda=1$ with a doubled value of $P$. If
two KIE's calculated with $\Delta\lambda=1$ at the two different values of $P
$ differed by a value that was lower then the sum of their statistical errors,
the lower value of $P $ was deemed sufficient for the calculation. The
statistical errors, i.e., root mean square errors (RMSE) were estimated with
the \textquotedblleft block-averaging\textquotedblright%
\ method\cite{Flyvbjerg_Petersen:1989} in order to remove the effect of
correlation length of the random walk in the Metropolis MC simulation.

In Ref. \onlinecite{Vanicek_Aoiz:2005}, $\eta_{\gamma}^{\ddagger}$ were taken
to be $0$ for all temperatures and all values of $\lambda$. Even though this
choice of DS positions leads to $C_{\mathrm{dd}}$ being stationary, it is a
local minimum rather than a saddle point. We therefore also checked the result
for the case when the proper optimal DS positions are found. Since from
symmetry considerations the optimal DS parameters satisfy $\eta_{a}^{\ddagger
}=-\eta_{b}^{\ddagger}$, simple bisection was sufficient to calculate the
values up to $0.01$ a.u. The results are presented in Table
\ref{tab:reproduced_h3}. Intermediate results of the calculations are
presented separately in Table \ref{tab:h3_intermediate} in Appendix C. We can
see that the values obtained with $\eta_{\gamma}^{\ddagger}=0$ agree well with
those of Ref. \onlinecite{Vanicek_Aoiz:2005}, validating our
modifications. It can also be seen that the full DS optimization improves
agreement of the QI results with the exact quantum result, making the method remarkably accurate
at low temperatures.

\begin{table}
\caption{Kinetic isotope effect $\mathrm{\cdot H+H_{2}/\cdot D+D_{2}}$ at different
temperatures.\hfill}\label{tab:reproduced_h3}
\begin{ruledtabular}
\begin{tabular}{cddD{.}{.}{2.7}D{.}{.}{2.7}D{.}{.}{2.2}D{,}{}{3.1}D{.}{.}{2.2}}
\multirow{2}{*}{$ T $ (K)} & \multicolumn{2}{c}{optimal $\eta^{\ddagger}_{a}=-\eta^{\ddagger}_{b}$} & \multicolumn{2}{c}{QI} & \multicolumn{1}{c}{\multirow{2}{*}{QM\footnotemark[1]}}
& \multicolumn{1}{c}{\multirow{2}{*}{\% error\footnotemark[2]}} & \multicolumn{1}{c}{\multirow{2}{*}{QI\footnotemark[1]}} \tabularnewline
\cline{2-3} \cline{4-5}
& \multicolumn{1}{c}{$\lambda=0$} & \multicolumn{1}{c}{$\lambda=1$} & \multicolumn{1}{c}{no DS optimization} & \multicolumn{1}{c}{optimized DS} & & & \tabularnewline
\hline
200 & 1.00 & 0.41 & 22.3\pm 0.2 & 22.6\pm 0.3 & 22.53 & <1 & 23.15 \tabularnewline
250 & 0.62 & 0 & 10.91\pm 0.08 & 9.92\pm 0.09  & 10.40 & -5 & 10.98 \tabularnewline
300 & 0.01 & 0 & 7.38\pm 0.05 & 7.35\pm 0.05  & 6.97 & 5 & 7.41 \tabularnewline
400 & 0 & 0 & \multicolumn{1}{c}{\_} & 4.87\pm 0.03 & 4.74 & 3 & 4.84 \tabularnewline
600 & 0 & 0 & \multicolumn{1}{c}{\_} & 3.29\pm 0.02 & 3.42 & -4 & 3.25 \tabularnewline
1000 & 0 & 0 & \multicolumn{1}{c}{\_} & 2.23\pm 0.01 & 2.61 & -15 & 2.22 \tabularnewline
1500 & 0 & 0 & \multicolumn{1}{c}{\_} & 1.81\pm 0.01 & 2.27 & -20 & 1.83 \tabularnewline
2400 & 0 & 0 & \multicolumn{1}{c}{\_} & 1.55\pm 0.01 & \multicolumn{1}{c}{\_} & \multicolumn{1}{c}{\_} & 1.56
\end{tabular}
\footnotetext[1]{Ref.~\onlinecite{Vanicek_Aoiz:2005}. QM denotes exact quantum-mechanical results from this reference.}
\footnotetext[2]{The error is defined as $ (\mathrm{KIE_{QI}}-\mathrm{KIE_{QM}})/\mathrm{KIE_{QM}}\times 100 \% $ for the optimized DS case.}
\end{ruledtabular}
\end{table}

\subsection{$\mathbf{CH_{4}+\cdot H\rightleftharpoons\cdot
CH_{3}+H_{2}}$ exchange}

As mentioned, the KIE's on the $\mathrm{CH_{4}+\cdot H\rightleftharpoons\cdot
CH_{3}+H_{2}}$ exchange had been studied by various numerical methods, but not
by the QI approximation. We therefore decided to test the accelerated QI\ method
on this reaction using the potential energy surface published in Ref. \onlinecite{Zhang_Braams_Bowman:2006}.

\subsubsection{Computational details}

We first ran a series of trial simulations to roughly determine the value of $P $
and the number of MC steps ensuring that at the lowest temperature the relative
statistical error of the KIE is below $1\%$
and that the discretization error with respect to $P $ is even smaller.
The target statistical error was guaranteed by running
$6.4\times10^{7}$ step MC simulations at $\lambda=0$ and $\lambda=1$, and
$3.2\times10^{7}$ simulations at other values of $\lambda$. The target
discretization error was achieved with $P=80 $ for the LT
and $P=20 $ for the combination of fourth-order splittings at $T=400\,\textmd{K}$; at
other temperatures $P $ was chosen such that the ratio $\beta/P$ stayed
approximately constant. We chose $\Delta\lambda=0.1$ as for the case of
$\mathrm{\cdot H+H_{2}/\cdot D + D_{2}}$; to be completely sure that the thermodynamic
integration error was negligible to the statistical one, we also ran
calculations with $\Delta\lambda=0.05$ at $T=400 \mathrm{K}$ for the equilibrium isotope effect
$\mathrm{\cdot CH_{3}/\cdot CD_{3}} $ and KIE $\mathrm{\cdot CH_{3}+D_{2}}/\mathrm{\cdot CD_{3}+D_{2}}$,
as these cases
exhibited the most drastic changes of properties during thermodynamic integration.

To determine the stationary positions of the DS's ($\{\eta
_{\gamma}^{\ddagger}\}$) we ran several short ($8\times10^{6}$ steps)
simulations to find the sign of $\partial\ln C_{\mathrm{dd}}/\partial
\eta_{\gamma}^{\ddagger}$ at different DS positions; the saddle points were
found with accuracy of $0.01$ a.u. The difference between $\eta_{a}^{\ddagger
}$ and $\eta_{b}^{\ddagger}$ turned out to be negligible at all temperatures
considered, in accordance with what is expected at \textquotedblleft
high\textquotedblright\ temperatures.\cite{Zhao_Miller:2004} The calculated
values of $\eta^{\ddagger}$ are presented in Tab.
\ref{tab:Optimal-DS-positions} in Appendix C; as expected, they are quite close
to the position of the classical transition state at $\eta^{\ddagger}=-0.94$.

\subsubsection{Results}

Next, we compared results obtained by the accelerated method (employing a
combination of fourth-order splittings and virial estimators) and by the
standard method (employing a combination of LT splitting and
thermodynamic estimators). The corresponding numerical results are labeled as
\textquotedblleft accel.\textquotedblright\ and \textquotedblleft
std.,\textquotedblright\ respectively. For further comparison, we calculated
the same KIE's also with the conventional transition state theory
(TST)\cite{Eyring:1935,Evans_Polanyi:1935,Mahan:1974} and TST with Wigner
tunneling correction\cite{Wigner:1932} (in the tables the corresponding
columns are denoted as \textquotedblleft TST\textquotedblright\ and
\textquotedblleft TST + Wigner\textquotedblright\ respectively). In the TST
framework the expression for the rate constant takes the form
\begin{equation}
k_{\mathrm{TST}}=\frac{k_{\mathrm{B}}T}{h}\frac{Q^{\ddagger}}{Q_{r}},
\label{eq:conventional_TST}%
\end{equation}
where $Q^{\ddagger}$ and $Q_{r}$ are partition functions of the transition and
reactant states, computed assuming separability of rotations and vibrations,
harmonic approximation for vibrations, and rigid rotor approximation for
rotations. Note that the usual factor $\exp(-E_{a}/k_{B}T) $, where $E_{a}$ is
the activation energy, is absorbed into our definition of $Q^{\ddagger}$ since
we use the same zero of energy for both $Q^{\ddagger}$ and $Q_{r}$. This
expression can be multiplied by the so-called Wigner tunneling correction
\begin{equation}
\kappa=1+\frac{1}{24}(\hbar\beta|\omega^{\ddagger}|)^{2}
\label{eq:Wigner_tun_cor}%
\end{equation}
to account for tunneling contribution to the reaction rate. Here
$\omega^{\ddagger}$ is the imaginary frequency corresponding to the motion
along the reaction coordinate,
\begin{equation}
\omega^{\ddagger}=\sqrt{\frac{K^{\ddagger}}{\mu^{\ddagger}}},
\end{equation}
$\mu^{\ddagger}$ is the effective reduced mass of the movement along the
reaction coordinate at the saddle point, and $K^{\ddagger} $ is the
corresponding negative force constant. Since the conventional TST expression
captures the changes of zero point energy as well as of the rotational and
translational partition functions due to the isotopic substitution, one may
expect that the difference between the QI and conventional TST should be
largely due to the difference between the extent of tunneling present in the
two isotopologs. The results are presented in Tables \ref{tab:ch5_ch4d}%
-\ref{tab:ch3h2_ch3d2}.

\begin{table}
[ptb]%
\caption{Kinetic isotope effect $\mathrm{CH_{4}+\cdot H}/\mathrm{CH_{4}+\cdot D}$. \hfill}\label{tab:ch5_ch4d}
\begin{ruledtabular}
\begin{tabular}{cccccccccc}
\multirow{2}{*}{$ T $ (K)}& \multirow{2}{*}{TST} & \multirow{2}{*}{\begin{tabular}[c]{@{}c@{}}TST\\+Wigner\end{tabular}} & \multicolumn{2}{c}{QI} & \multirow{2}{*}{TST\footnotemark[1]} & \multirow{2}{*}{CVT/$\mathrm{\mu OMT}$\footnotemark[1]} & \multirow{2}{*}{RDQD\footnotemark[2]} & \multirow{2}{*}{RPMD\footnotemark[3]} & \multirow{2}{*}{Expt.\footnotemark[4]}\tabularnewline
\cline{4-5}
& & &  accel. & std. &  &  &  &  & \tabularnewline
\hline
400 & 0.56 & 0.56 & 0.60$\pm 0.01$ & 0.62$\pm 0.07$ & 0.54 & 0.58 & 0.64 &  & 0.74\footnotemark[5]\tabularnewline
500 & 0.66 & 0.66 & 0.70$\pm 0.01$ & 0.65$\pm 0.08$ & 0.65 & 0.67 & 1.03 & 0.65 & 0.84\footnotemark[5]\tabularnewline
600 & 0.74 & 0.74 & 0.78$\pm 0.01$ & 0.7$\pm 0.1$ & 0.73 & 0.74 & 1.23 &  & 0.91\footnotemark[5]\tabularnewline
700 & 0.79 & 0.79 & 0.84$\pm 0.01$ & 0.9$\pm 0.1$ & 0.78 & 0.79 & 1.33 & 0.80 & 0.97\footnotemark[5]
\end{tabular}
\footnotetext[1]{Ref. \onlinecite{Pu_Truhlar:2002}}
\footnotetext[2]{Ref. \onlinecite{Kerkeni_Clary:2004}}
\footnotetext[3]{Ref. \onlinecite{Li_Guo:2013}}
\footnotetext[4]{Ref. \onlinecite{Kurylo_Timmons:1970}}
\footnotetext[5]{Values taken from Ref. \onlinecite{Pu_Truhlar:2002}}
\end{ruledtabular}

\end{table}

\begin{table}
[ptb]%
\caption{Kinetic isotope effect $\mathrm{\cdot CH_{3}+D_{2}}/\mathrm{\cdot CD_{3}+D_{2}}$. \hfill}\label{tab:ch3d2_cd3d2}
\begin{ruledtabular}
\begin{tabular}{cccccccc}
\multirow{2}{*}{$ T $ (K)} & \multirow{2}{*}{TST} & \multirow{2}{*}{TST+Wigner} & \multicolumn{2}{c}{QI} & \multirow{2}{*}{TST\footnotemark[1]} & \multirow{2}{*}{CVT/$\mathrm{\mu OMT}$\footnotemark[1]} & \multirow{2}{*}{Expt.\footnotemark[2]}\tabularnewline
\cline{4-5}
&  &  & accel. & std. & & & \tabularnewline
\hline
400 & 0.73 & 0.74 & 0.76$\pm 0.01$ & 0.7$\pm 0.1$ & 0.75 & 0.74 & 0.59\footnotemark[3]\tabularnewline
500 & 0.82 & 0.82 & 0.83$\pm 0.01$ & 0.9$\pm 0.1$ & 0.83 & 0.82 & 0.72\footnotemark[3]\tabularnewline
600 & 0.87 & 0.88 & 0.88$\pm 0.01$ & 0.8$\pm 0.2$ & 0.88 & 0.88 & 0.82\footnotemark[3]\tabularnewline
700 & 0.91 & 0.91 & 0.90$\pm 0.01$ & 0.8$\pm 0.2$ & 0.92 & 0.91 & 0.90\footnotemark[3]
\end{tabular}
\footnotetext[1]{Ref. \onlinecite{Pu_Truhlar:2002}}
\footnotetext[2]{Based on data from Refs. \onlinecite{Shapiro_Weston:1972,Tsang_Hampson:1986,Kerr_Parsonage:1976}}
\footnotetext[3]{Values taken from Ref. \onlinecite{Truhlar_Lynch:1992}}
\end{ruledtabular}

\end{table}

\begin{table}
[ptb]%
\caption{Kinetic isotope effect $\mathrm{\cdot CH_{3}+H_{2}}/\mathrm{\cdot CD_{3}+H_{2}}$.\hfill}\label{tab:ch3h2_cd3h2}
\begin{ruledtabular}
\begin{tabular}{cccccccc}
\multirow{2}{*}{$ T $ (K)} & \multirow{2}{*}{TST} & \multirow{2}{*}{TST+Wigner} & \multicolumn{2}{c}{QI} & \multirow{2}{*}{TST\footnotemark[1]} & \multirow{2}{*}{CVT/$\mathrm{\mu OMT}$\footnotemark[1]} & \multirow{2}{*}{Expt.\footnotemark[2]}\tabularnewline
\cline{4-5}
& & & accel. & std. & & & \tabularnewline
\hline
400 & 0.74 & 0.74 & 0.80$\pm 0.01$ & 0.78$\pm 0.08$ & 0.75 & 0.81 & 0.85\footnotemark[3]\tabularnewline
500 & 0.82 & 0.83 & 0.86$\pm 0.01$ & 0.9$\pm 0.1$ & 0.83 & 0.88 & 0.86\footnotemark[3]\tabularnewline
600 & 0.87 & 0.88 & 0.90$\pm 0.01$ & 1.0$\pm 0.2$ & 0.88 & 0.92 & 0.87\footnotemark[3]\tabularnewline
700 & 0.91 & 0.91 & 0.92$\pm 0.01$ & 1.0$\pm 0.2$ & 0.92 & 0.95 & 0.88\footnotemark[3]
\end{tabular}
\footnotetext[1]{Ref. \onlinecite{Pu_Truhlar:2002}}
\footnotetext[2]{Ref. \onlinecite{Shapiro_Weston:1972}}
\footnotetext[3]{Values taken from Ref. \onlinecite{Espinosa-Garcia_Corchado:1996}}
\end{ruledtabular}

\end{table}

\begin{table}
[ptb]%
\caption{Kinetic isotope effect $\mathrm{\cdot CH_{3}+HD}/\mathrm{\cdot CH_{3}+DH}$. \hfill}\label{tab:ch3hd_ch3dh}
\begin{ruledtabular}
\begin{tabular}{cccccccc}
\multirow{2}{*}{$ T $ (K)} & \multirow{2}{*}{TST} & \multirow{2}{*}{TST+Wigner} & \multicolumn{2}{c}{QI} & \multirow{2}{*}{TST\footnotemark[1]} & \multirow{2}{*}{CVT/$\mathrm{\mu OMT}$\footnotemark[1]} & \multirow{2}{*}{Expt.\footnotemark[2]}\tabularnewline
\cline{4-5}
&  &  & accel. & std. &  &  & \tabularnewline
\hline
467 & 1.51 & 1.86 & 2.10$\pm 0.02$ & 2.5$\pm 0.4$ & 1.50 & 1.83 & 2.1$\pm$0.5\footnotemark[3]\tabularnewline
531 & 1.48 & 1.76 & 1.84$\pm 0.02$ & 2.0$\pm 0.3$ & 1.47 & 1.71 & 1.9$\pm$0.3\footnotemark[3]\tabularnewline
650 & 1.44 & 1.64 & 1.59$\pm 0.02$ & 1.4$\pm 0.3$ & 1.43 & 1.56 & 1.2$\pm$0.3\footnotemark[3]\tabularnewline
\end{tabular}
\footnotetext[1]{Ref. \onlinecite{Pu_Truhlar:2002}}
\footnotetext[2]{Ref. \onlinecite{Shapiro_Weston:1972}}
\footnotetext[3]{Values taken from Ref. \onlinecite{Espinosa-Garcia_Corchado:1996}}
\end{ruledtabular}

\end{table}

\begin{table}
[ptb]%
\caption{Kinetic isotope effect $\mathrm{\cdot CD_{3}+HD}/\mathrm{\cdot CD_{3}+DH}$. \hfill}\label{tab:cd3hd_cd3dh}
\begin{ruledtabular}
\begin{tabular}{cccccccc}
\multirow{2}{*}{$ T $ (K)} & \multirow{2}{*}{TST} & \multirow{2}{*}{TST+Wigner} & \multicolumn{2}{c}{QI} & \multirow{2}{*}{TST\footnotemark[1]} & \multirow{2}{*}{CVT/$\mathrm{\mu OMT}$\footnotemark[1]} & \multirow{2}{*}{Expt.\footnotemark[2]}\tabularnewline
\cline{4-5}
&  &  & accel. & std. &  &  & \tabularnewline
\hline
400 & 1.56 & 1.99 & 2.52$\pm 0.02$ & 2.3$\pm 0.3$ & 1.55 & 1.91 & 1.85\footnotemark[3]\tabularnewline
500 & 1.50 & 1.80 & 1.95$\pm 0.02$ & 1.9$\pm 0.3$ & 1.49 & 1.60 & 1.61\footnotemark[3]\tabularnewline
600 & 1.46 & 1.68 & 1.65$\pm 0.02$ & 1.3$\pm 0.3$ & 1.45 & 1.56 & 1.47\footnotemark[3]\tabularnewline
700 & 1.43 & 1.60 & 1.52$\pm 0.01$ & 1.6$\pm 0.3$ & 1.42 & 1.49 & 1.38\footnotemark[3]
\end{tabular}
\footnotetext[1]{Ref. \onlinecite{Pu_Truhlar:2002}}
\footnotetext[2]{Ref. \onlinecite{Shapiro_Weston:1972}}
\footnotetext[3]{Values taken from Ref. \onlinecite{Pu_Truhlar:2002}}
\end{ruledtabular}

\end{table}

\begin{table}
[ptb]%
\caption{Kinetic isotope effect $\mathrm{\cdot CD_{3}+H_{2}}/\mathrm{\cdot CD_{3}+D_{2}}$. \hfill}\label{tab:cd3h2_cd3d2}
\begin{ruledtabular}
\begin{tabular}{cccccccc}
\multirow{2}{*}{$ T $ (K)} & \multirow{2}{*}{TST} & \multirow{2}{*}{TST+Wigner} & \multicolumn{2}{c}{QI} & \multirow{2}{*}{TST\footnotemark[1]} & \multirow{2}{*}{CVT/$\mathrm{\mu OMT}$\footnotemark[1]} & \multirow{2}{*}{Expt.\footnotemark[2]}\tabularnewline
\cline{4-5}
&  &  & accel. & std. &  &  & \tabularnewline
\hline
400 & 3.45 & 4.39 & 5.60$\pm 0.04$ & 5.0$\pm 0.8$ & 3.22 & 4.13 & 3.33\footnotemark[3]\tabularnewline
500 & 2.98 & 3.57 & 3.92$\pm 0.03$ & 3.9$\pm 0.5$ & 2.83 & 3.21 & 2.88\footnotemark[3]\tabularnewline
600 & 2.64 & 3.04 & 3.15$\pm 0.03$ & 2.6$\pm 0.6$ & 2.54 & 2.73 & 2.61\footnotemark[3]\tabularnewline
700 & 2.40 & 2.68 & 2.75$\pm 0.02$ & 2.5$\pm 0.4$ & 2.33 & 2.43 & 2.43\footnotemark[3]
\end{tabular}
\footnotetext[1]{Ref. \onlinecite{Pu_Truhlar:2002}}
\footnotetext[2]{Ref. \onlinecite{Shapiro_Weston:1972}}
\footnotetext[3]{Values taken from Ref. \onlinecite{Pu_Truhlar:2002}}
\end{ruledtabular}

\end{table}

\begin{table}
[ptb]%
\caption{Kinetic isotope effect $\mathrm{\cdot CH_{3}+H_{2}}/\mathrm{\cdot CH_{3}+D_{2}} $. \hfill}\label{tab:ch3h2_ch3d2}
\begin{ruledtabular}
\begin{tabular}{cccccccc}
\multirow{2}{*}{$ T $ (K)} & \multirow{2}{*}{TST} & \multirow{2}{*}{TST+Wigner} & \multicolumn{2}{c}{QI} & \multirow{2}{*}{TST\footnotemark[1]} & \multirow{2}{*}{CVT/$\mathrm{\mu OMT}$ \footnotemark[1]} & \multirow{2}{*}{Expt.\footnotemark[2]}\tabularnewline
\cline{4-5}
& & & accel. & std. & & & \tabularnewline
\hline
400 & 3.45 & 4.41 & 5.93$\pm 0.05$ & 5.8$\pm 0.8$ & 3.22 & 4.57 & 4.8$\pm$0.4\footnotemark[3]\tabularnewline
500 & 2.97 & 3.58 & 4.09$\pm 0.04$ & 4.0$\pm 0.6$ & 2.83 & 3.43 & 3.5$\pm$0.2\footnotemark[3]\tabularnewline
600 & 2.64 & 3.05 & 3.21$\pm 0.03$ & 3.1$\pm 0.5$ & 2.54 & 2.86 & 2.8$\pm$0.2\footnotemark[3]
\end{tabular}
\footnotetext[1]{Ref. \onlinecite{Pu_Truhlar:2002}}
\footnotetext[2]{Ref. \onlinecite{Shapiro_Weston:1972}}
\footnotetext[3]{Values taken from Ref. \onlinecite{Espinosa-Garcia_Corchado:1996}}
\end{ruledtabular}

\end{table}

First of all, it can be seen that for KIE's due to mass changes not affecting
the transferred atom (see Tables \ref{tab:ch5_ch4d}-\ref{tab:ch3h2_cd3h2}) the
QI values are close to those obtained by conventional TST. This can be
understood qualitatively from the expression (\ref{eq:Wigner_tun_cor}) for
Wigner tunneling correction for reaction rates. The main contribution to
$\mu^{\ddagger}$ appearing in the expression for $\omega^{\ddagger}$ comes
from the transferred atom, therefore if its mass does not change, the Wigner
tunneling corrections for different isotopologs will have similar values and
largely cancel out in the KIE.

Second, note that, in agreement with the usual difference in magnitudes of
secondary and primary isotope effects, replacing $\mathrm{\cdot CH_{3}}$ with
$\mathrm{\cdot CD_{3}}$ leads to a much smaller rate change than does
replacing $\mathrm{H_{2}}$ with $\mathrm{D_{2}}$ (compare Tables
\ref{tab:ch3d2_cd3d2}-\ref{tab:ch3h2_cd3h2} and \ref{tab:cd3h2_cd3d2}%
-\ref{tab:ch3h2_ch3d2}) This consideration also explains why the KIE's
corresponding to $\mathrm{\cdot CH_{3}+H_{2}}/\mathrm{\cdot CH_{3}+D_{2}}$ and
$\mathrm{\cdot CD_{3}+H_{2}}/\mathrm{\cdot CD_{3}+D_{2}}$ are quite close to
each other (see Tables \ref{tab:ch3h2_ch3d2} and \ref{tab:cd3h2_cd3d2}).
For some KIE's presented in Tables \ref{tab:cd3hd_cd3dh}-\ref{tab:ch3h2_ch3d2}
it appears that results obtained with TST or TST with
Wigner tunneling correction are in better agreement with experimental values
than those obtained with the QI, probably indicating that
a large cancellation takes place between the errors of the TST and of the potential energy surface (PES).

In order to estimate the influence of the used force field on the final result
we also ran calculations with the PES published in Ref.
\onlinecite{Corchado_Espinosa-Garcia:2009} for $\mathrm{CH_{4}+\cdot
H/CH_{4}+\cdot D}$. After finding the optimal DS positions (see Table
\ref{tab:Optimal-DS-positions} in Appendix C), we compared the QI values of
this KIE obtained with the two PES's from
Refs.~\onlinecite{Corchado_Espinosa-Garcia:2009} and
\onlinecite{Zhang_Braams_Bowman:2006} (see Table \ref{tab:ch5_ch4d_eg}),
finding that the choice of the PES affects the KIE value by as much as $10\%$.
In contrast, comparison of the KIE's computed with the same PES, but with two
different accurate quantum methodologies (RPMD and QI) results in a remarkable
agreement, within the statistical error of less than $2\%$. Finally, note
that the QI\ KIE is in much better agreement with
experiment if computed with the PES of
Ref.~\onlinecite{Zhang_Braams_Bowman:2006} than with the PES of
Ref.~\onlinecite{Corchado_Espinosa-Garcia:2009}, suggesting that the former
PES, which was used for most of the calculations in this paper, was the
appropriate choice.

\begin{table}
[ptb]%
\caption{Influence of the potential energy surface (PES) on the KIE $\mathrm{CH_{4}+\cdot H}/\mathrm{CH_{4}+\cdot D}$.
Comparison of the QI KIE's calculated using the PES's of Refs. \onlinecite{Corchado_Espinosa-Garcia:2009}
and \onlinecite{Zhang_Braams_Bowman:2006}. Note also the remarkable agreement
between the KIE's computed with RPMD and QI on the same PES.
\hfill}\label{tab:ch5_ch4d_eg} \begin{ruledtabular}
\begin{tabular}{ccccccc}
\multirow{2}{*}{$ T $ (K)} & \multicolumn{4}{c}{PES of Ref. \onlinecite{Corchado_Espinosa-Garcia:2009}} & PES of Ref. \onlinecite{Zhang_Braams_Bowman:2006}&
\multirow{2}{*}{Expt.\footnotemark[2]} \tabularnewline
\cline{2-5} \cline{6-6}
& TST & TST+Wigner & RPMD\footnotemark[1] & QI & QI & \tabularnewline
\hline
400 & 0.52 & 0.52 &  & 0.54$\pm 0.01$ & 0.60$\pm 0.01$ & 0.74 \footnotemark[3]\tabularnewline
500 & 0.63 & 0.63 & 0.65 & 0.64$\pm 0.01$ & 0.70$\pm 0.01$ & 0.84 \footnotemark[3]\tabularnewline
600 & 0.71 & 0.71 &  & 0.73$\pm 0.01$ & 0.78$\pm 0.01$ & 0.91 \footnotemark[3]\tabularnewline
700 & 0.77 & 0.77 & 0.80 & 0.79$\pm 0.01$ & 0.84$\pm 0.01$ & 0.97 \footnotemark[3]
\end{tabular}
\footnotetext[1]{Ref. \onlinecite{Li_Guo:2013}}
\footnotetext[2]{Ref. \onlinecite{Kurylo_Timmons:1970}}
\footnotetext[3]{Values taken from Ref. \onlinecite{Pu_Truhlar:2002}}
\end{ruledtabular}

\end{table}

As for the performance of the
fourth-order splittings, since an analytical gradient was not available for 
the $ \mathrm{CH}_{4}+\cdot\mathrm{H}$ system, 
the gradient had to be calculated numerically using finite differences.
For constrained simulations this made the force twelve times (once per each internal degree of freedom) as expensive as
the potential itself, leading to a seven-fold increase in CPU time for a given $ P $ and 
number of MC steps when the
SC splitting was used. Since the fourth-order splitting
decreased the necessary $ P$ by a factor of four, the final increase in CPU time for
a given discretization error and number of MC steps was $ 75\%$.
For unconstrained simulations employed to find
$ \cdot\mathrm{CH}_{3}/\cdot\mathrm{CD}_{3}$ equilibrium isotope effect the force was six times as expensive as the potential;
since the use of the TI factorization allowed to decrease $ P$ four times, 
the final increase in CPU time was also $ 75\%$
for a given number of MC steps and discretization error.

In summary, the KIE's were reproduced in a reasonable agreement with
experiment. The differences are probably due to both the error of the
potential energy surface used and the large experimental error. 
Note that our
accelerated methodology again drastically reduced both the discretization and
statistical errors of the calculations.

\section{Conclusions\label{sec:conclusions}}

In conclusion, we have accelerated the methodology from Ref.
\onlinecite{Vanicek_Aoiz:2005} for computing KIE's with the QI\ approximation.
In particular, we have combined virial estimators (several of which have been
derived for the first time here) with high-order factorizations of the quantum
Boltzmann operator, and shown that this combination significantly
accelerates the QI calculations of the KIE's in systems with prominent
quantum effects. We have also proposed and demonstrated the utility of a new
method for the thermodynamic integration of the delta-delta correlation
function $C_{\text{dd}}$, which is a convenient alternative to the approach
employed in Ref. \onlinecite{Vanicek_Aoiz:2005}. Our accelerated methodology
has been tested on the $\mathrm{CH_{4}+\cdot H\rightleftharpoons\cdot
CH_{3}+H_{2}}$ model exchange, obtaining results that agree reasonably well
with published experimental values.

\begin{acknowledgments}
This research was supported by the Swiss National Science Foundation with
Grant No. 200020\_150098 and by the EPFL.
\end{acknowledgments}

\section*{Appendix A: derivation of the fourth-order corrections for different
estimators}

When one of the fourth-order factorizations is used, $V_{\mathrm{eff}}%
^{(s)}(\mathbf{r}^{(s)})$ has an explicit dependence on mass and $\beta$; as a
result one needs to add appropriate \textquotedblleft
corrections\textquotedblright\ to the estimators arising from the
differentiation with respect to these quantities.

For $\partial\mathrm{ln}Q_{r}/\partial\lambda/\beta$ and $\partial
\mathrm{ln}C_{\mathrm{dd}}/\partial\lambda/\beta$, it follows from Eqs.
\eqref{eq:F} and \eqref{eq:F_ts} that the correction $F_{r,\mathrm{grad}}$
is%
\begin{equation}
F_{r,\mathrm{grad}}=-\frac{\beta}{P^{3}}\sum_{i=1}^{N}\frac{\mathit{d}m_{i}%
}{\mathit{d}\lambda}\sum_{s=1}^{P}w_{s}d_{s}\frac{\partial V_{\mathrm{grad}%
}(\mathbf{r}^{(s)})}{\partial m_{i}}=\frac{\hbar^{2}\beta}{P^{3}}\sum
_{s=1}^{P}w_{s}d_{s}\sum_{i=1}^{N}\frac{1}{m_{i}^{2}}\frac{\mathit{d}m_{i}%
}{\mathit{d}\lambda}|\nabla_{i}V(\mathbf{r}^{(s)})|^{2}.
\end{equation}
Note that when a coordinate rescaling is used to obtain an estimator
(e.g., for centroid virial estimators), the correction remains the same
due to the following equality:%
\begin{equation}
\frac{\mathit{d}V_{\mathrm{eff}}^{(s)}[\mathbf{r}^{(s)}(m_{i}),m_{i}%
]}{\mathit{d}m_{i}}=\left\langle \frac{\partial\mathbf{r}^{(s)}}{\partial
m_{i}},\frac{\partial V_{\mathrm{eff}}^{(s)}}{\partial\mathbf{r}^{(s)}%
}\right\rangle _{0}+d_{s}\left(  \frac{\beta}{P}\right)  ^{2}\frac{\partial
V_{\mathrm{grad}}}{\partial m_{i}}.
\end{equation}

As for $F_{\mathrm{th}}$ and $F_{\mathrm{v}}$, since%
\begin{equation}
\frac{\partial V_{\mathrm{eff}}^{(s)}(\mathbf{r}^{(s)})}{\partial\beta}%
=2d_{s}\frac{\beta}{P^{2}}V_{\mathrm{grad}}(\mathbf{r}^{(s)})\text{,}%
\end{equation}
the gradient correction to be added is%
\begin{equation}
F_{\mathrm{grad}}=\frac{4\beta^{2}}{P^{3}}\left(  \sum_{s=1}^{P/2-1}%
-\sum_{s=P/2+1}^{P-1}\right)  w_{s}d_{s}V_{\mathrm{grad}}(\mathbf{r}%
^{(s)})\text{.}%
\end{equation}
Again, this correction is the same for the virial and thermodynamic variants.

Since the $G$ factor involves the second derivatives with respect to $\beta$,
the corrections will be different for $G_{\mathrm{th}}$ and $G_{\mathrm{v}}$.
While $G_{\mathrm{th,grad}}$ is obtained easily as%
\begin{equation}
G_{\mathrm{th,grad}}=-\frac{24\beta}{P^{3}}\sum_{s=1}^{P}w_{s}%
d_{s}V_{\mathrm{grad}}(\mathbf{r}^{(s)}),
\end{equation}
to find $G_{\mathrm{v}}$, one needs to take advantage of the following
relations:%
\begin{equation}
\frac{\mathit{d}V_{\mathrm{eff}}^{(s)}[\mathbf{r}^{(s)}(\beta),\beta
]}{\mathit{d}\beta}=\left\langle \frac{\partial\mathbf{r}^{(s)}(\beta
)}{\partial\beta},\frac{\partial V_{\mathrm{eff}}^{(s)}(\mathbf{r}^{(s)}%
)}{\partial\mathbf{r}^{(s)}}\right\rangle _{0}+\frac{\partial V_{\mathrm{eff}%
}^{(s)}(\mathbf{r}^{(s)})}{\partial\beta}\text{,}%
\end{equation}
\begin{widetext}
\begin{equation}
\begin{split}\frac{\mathit{d}^{2}V^{(s)}_{\mathrm{eff}}[\mathbf{r}^{(s)}(\beta),\beta]}{\mathit{d}\beta^{2}} =
& \left\langle\frac{\partial^{2}\mathbf{r}^{(s)}(\beta)}{\partial\beta^{2}},\frac{\partial V^{(s)}_{\mathrm{eff}}(\mathbf{r}^{(s)})}{\partial \mathbf{r}^{(s)}}\right\rangle_{0}
+\left\langle\frac{\partial\mathbf{r}^{(s)}(\beta)}{\partial\beta},\frac{\partial^{2}V^{(s)}_{\mathrm{eff}}(\mathbf{r}^{(s)})}{(\partial\mathbf{r}^{(s)})^{2}},
\frac{\partial\mathbf{r}^{(s)}(\beta)}{\partial\beta}\right\rangle_{00}\\
& +2\left\langle\frac{\partial\mathbf{r}^{(s)}(\beta)}{\partial\beta},\frac{\partial}{\partial\mathbf{r}^{(s)}}\left[\frac{\partial V^{(s)}_{\mathrm{eff}}(\mathbf{r}^{(s)})}{\partial\beta}\right]\right\rangle_{0}
+\frac{\partial^{2}V^{(s)}_{\mathrm{eff}}(\mathbf{r}^{(s)})}{\partial\beta^{2}} \text{,}
\end{split}
\end{equation}
\begin{equation}
\begin{split}
\frac{\mathit{d}^{2}\{\beta V^{(s)}_{\mathrm{eff}}[\mathbf{r}^{(s)}(\beta),\beta]\}}{\mathit{d}\beta^{2}}= & 2\frac{\mathit{d}V^{(s)}_{\mathrm{eff}}(\mathbf{r}^{(s)})}{\mathit{d}\beta}
+\beta\frac{\mathit{d}^{2}V^{(s)}_{\mathrm{eff}}(\mathbf{r}^{(s)})}{\mathit{d}\beta^{2}}\\
= & \left\langle\left(2\frac{\partial\mathbf{r}^{(s)}(\beta)}{\partial\beta}+\beta\frac{\partial^{2}\mathbf{r}^{(s)}(\beta)}{\partial\beta^{2}}\right),\frac{\partial V^{(s)}_{\mathrm{eff}}(\mathbf{r}^{(s)})}
{\partial\mathbf{r}^{(s)}}\right\rangle_{0}\\
& +\beta\left\langle\frac{\partial\mathbf{r}^{(s)}(\beta)}{\partial\beta},\frac{\partial^{2}V^{(s)}_{\mathrm{eff}}(\mathbf{r}^{(s)})}{(\partial\mathbf{r}^{(s)})^{2}},
\frac{\partial\mathbf{r}^{(s)}(\beta)}{\partial\beta}\right\rangle_{00} \\
& +2\beta\left\langle\frac{\partial\mathbf{r}^{(s)}(\beta)}{\partial\beta},
\frac{\partial}{\partial\mathbf{r}^{(s)}(\beta)}\left[\frac{\partial V^{(s)}_{\mathrm{eff}}(\mathbf{r}^{(s)})}{\partial\beta}\right]\right\rangle_{0}
+\beta\frac{\partial^{2}V^{(s)}_{\mathrm{eff}}(\mathbf{r}^{(s)})}{\partial\beta^{2}}+2\frac{\partial V^{(s)}_{\mathrm{eff}}(\mathbf{r}^{(s)})}{\partial\beta}\text{,}
\end{split}
\end{equation}
The only terms for which the explicit $\beta$ dependence plays a
role are the last three. As a result we get:
\begin{equation}
\begin{split}
G_{\mathrm{v,grad}} & =-\frac{4}{P}\sum_{s=1}^{P}w_{s}\left\{ 2\beta\left\langle\frac{\partial\mathbf{r}^{(s)}(\beta)}{\partial\beta},\frac{\partial}
{\partial\mathbf{r}^{(s)}(\beta)}\left[\frac{\partial V^{(s)}_{\mathrm{eff}}(\mathbf{r}^{(s)})}{\partial\beta}\right]\right\rangle_{0}
+\beta\frac{\partial^{2}V^{(s)}_{\mathrm{eff}}(\mathbf{r}^{(s)})}{\partial\beta^{2}}+2\frac{\partial V^{(s)}_{\mathrm{eff}}(\mathbf{r}^{(s)})}{\partial\beta}\right\}\text{.}
\end{split}
\end{equation}
\end{widetext}This expression can be rewritten as%
\begin{equation}
G_{\mathrm{v,grad}}=-\frac{8\beta}{P^{3}}\sum_{s=1}^{P}w_{s}%
d_{s}\left[  3V_{\mathrm{grad}}%
\vphantom{(\mathbf{r}^{(s)}-\check{\mathbf{r}}^{(s)})}\right.  \left.
+\left\langle (\mathbf{r}^{(s)}-\check{\mathbf{r}}^{(s)}),\nabla
V_{\mathrm{grad}}(\mathbf{r}^{(s)})\right\rangle _{0}\right]  .
\label{eq:G_v_corr}%
\end{equation}

\section*{Appendix B: derivation of $\mathbf{{B^{k(\gamma)}}}$}

The present derivation is a slight generalization of the one found in Ref.
\onlinecite{denOtter:2000}. We start by transforming to mass-scaled
coordinates,
\begin{equation}%
\begin{split}
\mathrm{x}_{i}^{(s)}  &  :=\sqrt{m_{i}}\mathrm{r}_{i}^{(s)},\\
\mathrm{x}_{\gamma i}  &  :=\sqrt{m_{i}}\mathrm{r}_{\gamma i},\\
\overline{\xi}_{\gamma}(\mathbf{x}_{\gamma})  &  :=\xi_{\gamma}(\mathbf{r}%
_{\gamma}),
\end{split}
\end{equation}
which will simplify the subsequent algebra due to the equality
\begin{equation}
||\nabla\xi_{\gamma}(\mathbf{r}_{\gamma})||_{-}=|\nabla\overline{\xi}_{\gamma
}(\mathbf{x}_{\gamma})|.
\end{equation}
In mass-scaled coordinates, the PI representation (\ref{eq:C_dd(PI)}) of the
delta-delta correlation function can be rewritten as
\begin{equation}
C_{\text{dd},P}=\int\overline{\rho}(\{\mathbf{x}^{(s)}\})\Delta\lbrack
\overline{\xi}_{\gamma}(\mathbf{x}_{\gamma})]\mathit{d}\{\mathbf{x}^{(s)}\},
\end{equation}
where the second normalized delta function has been absorbed into
$\overline{\rho}$ in order to simplify the following derivation.
Differentiation of $C_{\text{dd},P}$ with respect to the DS's
parameters yields
\begin{equation}%
\begin{split}
\frac{\partial C_{\text{dd},P}}{\partial\eta_{k}^{(\gamma)}}=  &
\frac{\partial}{\partial\eta_{k}^{(\gamma)}}\int\overline{\rho}(\{\mathbf{x}%
^{(s)}\})\Delta\left[  \overline{\xi}_{\gamma}(\mathbf{x}_{\gamma},\eta
_{k}^{(\gamma)})\right]\mathit{d}\{\mathbf{x}^{(s)}\}\\
=  &  \int\frac{\langle\nabla\overline{\xi}_{\gamma}(\mathbf{x}_{\gamma
}),\nabla\overline{\xi}_{\gamma}(\mathbf{x}_{\gamma})\rangle}{|\nabla
\overline{\xi}_{\gamma}(\mathbf{x}_{\gamma})|}\overline{\rho}(\{\mathbf{x}%
^{(s)}\})\frac{\partial\overline{\xi}_{\gamma}(\mathbf{x}_{\gamma})}%
{\partial\eta_{k}^{(\gamma)}}\frac{\mathit{d}}{\mathit{d}\overline{\xi
}_{\gamma}}\left\{  \delta\left[  \overline{\xi}_{\gamma}(\mathbf{x}_{\gamma
},\eta_{k}^{(\gamma)})\right]  \right\}  \mathit{d}\{\mathbf{x}^{(s)}\}\\
&  +\int\frac{\partial\mathrm{ln}|\nabla\overline{\xi}_{\gamma}(\mathbf{x}%
_{\gamma})|}{\partial\eta_{k}^{(\gamma)}}\overline{\rho}(\{\mathbf{x}%
^{(s)}\})\Delta\left[  \overline{\xi}_{\gamma}(\mathbf{x}_{\gamma},\eta
_{k}^{(\gamma)})\right]  \mathit{d}\{\mathbf{x}^{(s)}\}\\
=  &  \int\nabla\left\{  \delta\left[  \overline{\xi}_{\gamma}(\mathbf{x}%
_{\gamma},\eta_{k}^{(\gamma)})\right]  \right\}  \frac{\partial\overline{\xi
}_{\gamma}(\mathbf{x}_{\gamma})}{\partial\eta_{k}^{(\gamma)}}\frac
{\nabla\overline{\xi}_{\gamma}(\mathbf{x}_{\gamma})}{\left\vert \nabla
\overline{\xi}_{\gamma}(\mathbf{x}_{\gamma})\right\vert }\overline{\rho
}(\{\mathbf{x}^{(s)}\})\mathit{d}\{\mathbf{x}^{(s)}\}\\
&  +\int\frac{1}{|\nabla\overline{\xi}_{\gamma}(\mathbf{x}_{\gamma})|^{2}%
}\left\langle \nabla\overline{\xi}_{\gamma},\nabla\frac{\partial\overline{\xi
}_{\gamma}(\mathbf{x}_{\gamma})}{\partial\eta_{k}^{(\gamma)}}\right\rangle
\overline{\rho}(\{\mathbf{x}^{(s)}\})\Delta\left[  \overline{\xi}_{\gamma
}(\mathbf{x}_{\gamma},\eta_{k}^{(\gamma)})\right]  \mathit{d}\{\mathbf{x}%
^{(s)}\}.
\end{split}
\end{equation}
After integrating by parts with respect to $\mathbf{x}_{\gamma}$ in the first
integral, we get
\begin{equation}%
\begin{split}
\frac{\partial}{\partial\eta_{k}^{(\gamma)}}\int\overline{\rho}(\{\mathbf{x}%
^{(s)}\})\Delta\left[  \overline{\xi}_{\gamma}(\mathbf{x}_{\gamma},\eta
_{k}^{(\gamma)})\right]  \mathit{d}\{\mathbf{x}^{(s)}\}=  &  -\int%
\frac{\partial\overline{\xi}_{\gamma}(\mathbf{x}_{\gamma})}{\partial\eta
_{k}^{(\gamma)}}\left[  \left\langle \nabla\overline{\xi}_{\gamma}%
(\mathbf{x}_{\gamma}),\nabla^{(\gamma)}\mathrm{ln}\overline{\rho}%
(\{\mathbf{x}^{(s)}\})\right\rangle _{0}%
\vphantom{\left\langle\nabla,\frac{\nabla\overline{\xi}_{\gamma}(\mathbf{x}_{\gamma})} {|\nabla\overline{\xi}_{\gamma}(\mathbf{x}_{\gamma})|}\right\rangle_{0}}\right.
\\
&  \left.  +|\nabla\overline{\xi}_{\gamma}(\mathbf{x}_{\gamma})|\left\langle
\nabla,\frac{\nabla\overline{\xi}_{\gamma}(\mathbf{x}_{\gamma})}%
{|\nabla\overline{\xi}_{\gamma}(\mathbf{x}_{\gamma})|}\right\rangle
_{0}\right]  /|\nabla\overline{\xi}_{\gamma}(\mathbf{x}_{\gamma})|^{2}\\
&  \times\Delta\left[  \overline{\xi}_{\gamma}(\mathbf{x}_{\gamma},\eta
_{k}^{(\gamma)})\right]  \overline{\rho}(\{\mathbf{x}^{(s)}\})\mathit{d}%
\{\mathbf{x}^{(s)}\}\text{.}%
\end{split}
\end{equation}
Equation (\ref{eq:B_general}) is obtained by substituting the explicit
expression for $\overline{\rho}$ and transforming back to Cartesian coordinates.

\section*{Appendix C: Additional numerical results}

\begin{figure}
[ptbh]\includegraphics[width=\textwidth]{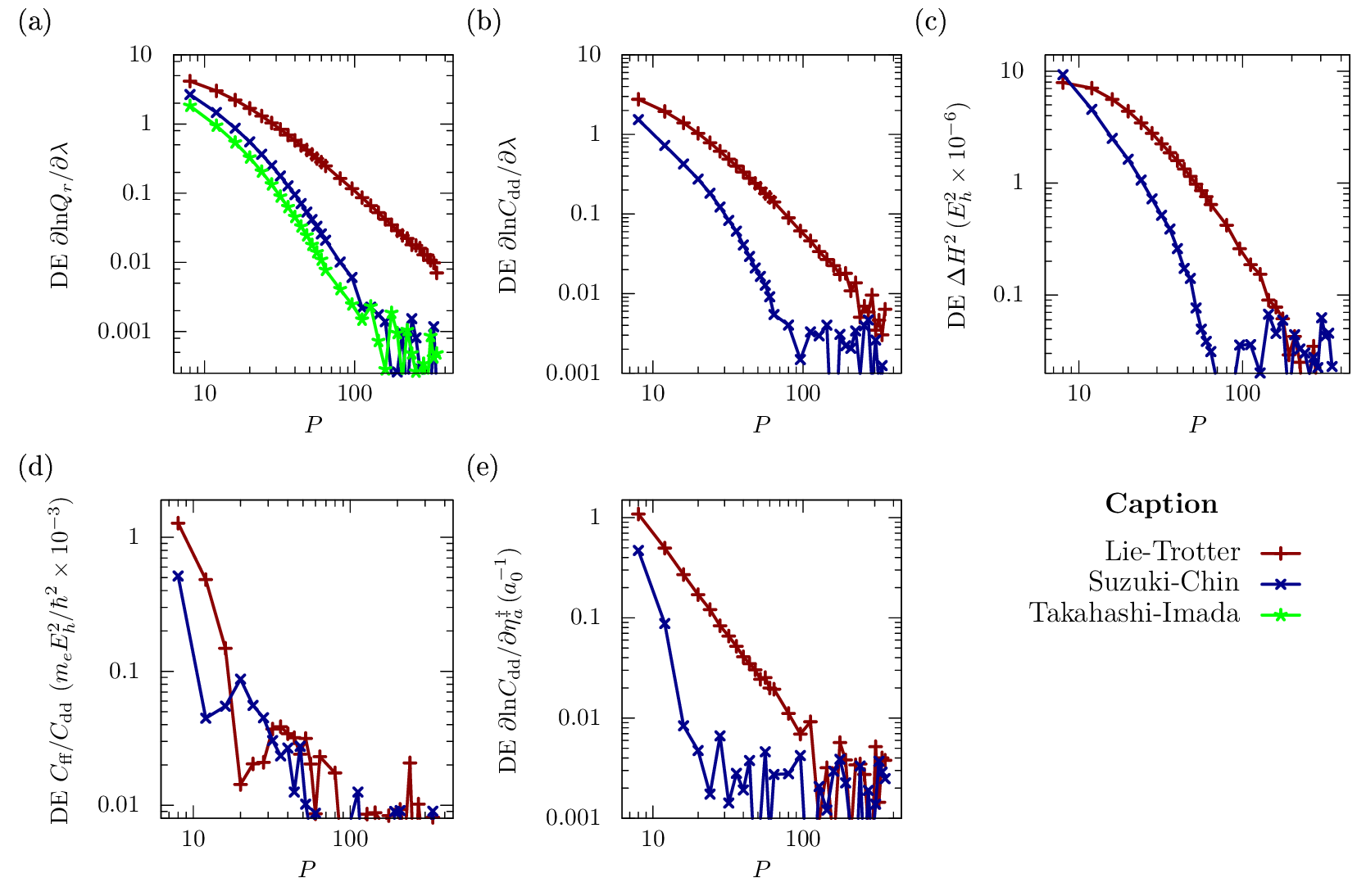}
\caption{Absolute discretization error (DE) of different quantities as a function of $ P $: (a) $\partial
\mathrm{ln} Q_{r}/\partial\lambda$, (b) $\partial\mathrm{ln}C_{\text{dd}}/\partial\lambda$, (c) $\Delta H^{2}$, (d)
$C_{\mathrm{ff}}/C_{\mathrm{dd}}$, (e)
$\partial\mathrm{ln} C_{\text{dd}}/\partial\eta_{a}^{\ddagger}$.
Results shown were obtained with the virial estimators and correspond to the KIE
$\mathrm{\cdot H+H_{2}/\cdot D+D_{2}}$ at 200 K.\hfill \label{fig:log_discr_err_convergence}}
\end{figure}

In this section we present some additional numerical results that were moved
from the main text for the sake of clarity. Figure
\ref{fig:log_discr_err_convergence} depicts the logarithmic plots of the
discretization errors of various ingredients of the QI\ approximation as
functions of the Trotter number $P$. The discretization error for a quantity
$A$ is defined as $|A_{P}-A_{\infty}|$, where $A_{\infty}$ was estimated by
averaging $A_{P}$ over several highest values of $P$, for which the
discretization error was considered negligible. The averaging was performed in
order to reduce the statistical error. The plots in
Fig.~\ref{fig:log_discr_err_convergence} demonstrate the faster convergence to
the quantum limit achieved with higher-order factorizations:\ indeed,
especially for the \ logarithmic derivative of $Q_{r}$, one can see that the
discretization error dependence approaches the asymptotic behavior
$\mathcal{O}(P^{-2})$ for the LT and $\mathcal{O}%
(P^{-4})$ for the SC and TI factorizations. In addition, in
all panels, it is clear for which value of $P$ the discretization error
becomes smaller than the statistical error, since for higher values of $P$ the
smooth dependence of the discretization error on $P$ is obscured by statistical noise.

Table~\ref{tab:h3_intermediate} contains values of various factors used to
obtain the results in Table~\ref{tab:reproduced_h3} for the QI KIE on the
reaction $\mathrm{\cdot H_{\alpha}+H_{\beta}H_{\gamma}\rightarrow H_{\alpha
}H_{\beta}+\cdot H_{\gamma}}$ with optimized DS. Finally,
Table~\ref{tab:Optimal-DS-positions} contains optimized DS positions that were
used for calculating KIE's on the $\mathrm{CH_{4}+\cdot H\rightleftharpoons
\cdot CH_{3}+H_{2}}$.

\begin{table}
\caption{Values of the factors entering the QI expression (\ref{eq:KIE_QI}) for the KIE
$\mathrm{\cdot H+H_{2}/\cdot D+D_{2}}$ with optimized dividing surface
positions, displayed in Table~\ref{tab:reproduced_h3}. All quantities as well as their statistical errors are in atomic units.\hfill}\label{tab:h3_intermediate}
\begin{ruledtabular}
\begin{tabular}{cD{.}{.}{3.7}D{.}{.}{3.7}D{,}{}{1.8}D{,}{}{1.8}D{,}{}{2.8}D{,}{}{4.8}}
\multirow{2}{*}{$ T $ (K)} & \multicolumn{2}{c}{$ \Delta H^{2} \times 10^{6}$} & \multicolumn{2}{c}{$C_{\mathrm{ff}}/C_{\mathrm{dd}} \times 10^{3}$} &
\multicolumn{1}{c}{\multirow{2}{*}{$ C_{\mathrm{dd}}$ ratio}} & \multicolumn{1}{c}{\multirow{2}{*}{$ Q_{r} $ ratio}}  \tabularnewline
\cline{2-3} \cline{4-5}
& \multicolumn{1}{c}{$\lambda=0$} & \multicolumn{1}{c}{$\lambda=1$} & \multicolumn{1}{c}{$\lambda=0$} & \multicolumn{1}{c}{$ \lambda=1$} & & \tabularnewline
\hline
200 &  3.68\pm 0.03 & 4.27\pm 0.02 & 1,.33\pm 0.01 & 2,.03 \pm 0.01 & 43,.8\pm 0.3 & 1404, \pm 1\tabularnewline
250 & 4.87\pm 0.04 & 4.95\pm 0.03 & 2,.01\pm 0.01 & 2,.10\pm 0.01 & 55,.8\pm 0.3 & 572,.9 \pm 0.3\tabularnewline
300 & 7.40\pm 0.04 & 5.10\pm 0.03 & 2,.73\pm 0.01 & 1,.97 \pm 0.01 & 49,.5\pm 0.1 & 316,.1 \pm 0.2 \tabularnewline
400 &  7.92\pm 0.05  & 6.27\pm 0.04  & 2,.49\pm 0.01 & 2,.01\pm 0.01 & 33,.96\pm 0.06 & 149,.5 \pm 0.1 \tabularnewline
600 &  12.0\pm 0.1 & 10.7 \pm 0.1  & 2,.77\pm 0.01 & 2,.44\pm 0.01 & 23,.11\pm 0.04 & 70,.92 \pm 0.03 \tabularnewline
1000 &  26.8\pm 0.1  & 24.5\pm 0.1  & 3,.82 \pm 0.03 & 3,.55 \pm 0.02 & 18,.15\pm 0.02 & 39,.42 \pm 0.01 \tabularnewline
1500 &  54.3\pm 0.2  & 50.3\pm 0.2  & 5,.29 \pm 0.04 & 5,.05 \pm 0.02 & 16,.78 \pm 0.02 & 30,.11 \pm 0.01 \tabularnewline
2400 &  124.6\pm 0.4 & 117.1\pm 0.4 & 7,.98\pm 0.05 & 7,.80\pm 0.03 & 16,.36 \pm 0.01 & 25,.51 \pm 0.01
\end{tabular}
\end{ruledtabular}
\end{table}

\begin{table}
[ptb]\caption{Optimal positions of the dividing surfaces along the reaction
coordinate [see Eq.~\eqref{eq:reaction_coordinate}] for transition states
of several isotopic variants of the $\mathrm{CH_{4}+\cdot H}\rightleftharpoons
\mathrm{\cdot CH_{3}+H_{2}}$ exchange at several temperatures. \hfill}\label{tab:Optimal-DS-positions}
\begin{ruledtabular}
\begin{tabular}{ccccc}
Potential energy surface of Ref. \onlinecite{Zhang_Braams_Bowman:2006}\tabularnewline
\hline
TS & 400 K & 500 K & 600 K & 700 K \tabularnewline
\hline
$\mathrm{H_{3}C\cdot\cdot\cdot H\cdot\cdot\cdot H}$ & -0.91 & -0.90 & -0.88 & -0.86\tabularnewline
$\mathrm{H_{3}C\cdot\cdot\cdot H\cdot\cdot\cdot D}$ & -0.88 & -0.87 & -0.85 & -0.84\tabularnewline
$\mathrm{D_{3}C\cdot\cdot\cdot H\cdot\cdot\cdot D}$ & -0.89 & -0.88 & -0.86 & -0.85\tabularnewline
$\mathrm{D_{3}C\cdot\cdot\cdot D\cdot\cdot\cdot H}$ & -0.93 & -0.91 & -0.89 & -0.86\tabularnewline
$\mathrm{H_{3}C\cdot\cdot\cdot D\cdot\cdot\cdot D}$ & -0.89 & -0.87 & -0.86 & -0.85\tabularnewline
$\mathrm{D_{3}C\cdot\cdot\cdot D\cdot\cdot\cdot D}$ & -0.90 & -0.89 & -0.87 & -0.85\tabularnewline
$\mathrm{D_{3}C\cdot\cdot\cdot H\cdot\cdot\cdot H}$ & -0.92 & -0.90 & -0.89 & -0.87\tabularnewline
\hline
& 467 K & 531 K & 650 K \tabularnewline
\hline
$\mathrm{H_{3}C\cdot\cdot\cdot H\cdot\cdot\cdot D}$ & -0.87 & -0.86 & -0.84\tabularnewline
$\mathrm{H_{3}C\cdot\cdot\cdot D\cdot\cdot\cdot H}$ & -0.90 & -0.89 & -0.87\tabularnewline
\hline
Potential energy surface of Ref. \onlinecite{Corchado_Espinosa-Garcia:2009}\tabularnewline
\hline
TS & 400 K & 500 K & 600 K & 700 K \tabularnewline
\hline
$\mathrm{H_{3}C\cdot\cdot\cdot H\cdot\cdot\cdot H}$ & -1.03 & -1.00 & -0.97 & -0.95\tabularnewline
$\mathrm{H_{3}C\cdot\cdot\cdot H\cdot\cdot\cdot D}$ & -1.00 & -0.97 & -0.94 & -0.92\tabularnewline
\end{tabular}%
\end{ruledtabular}

\end{table}

\FloatBarrier
\bibliographystyle{aipnum4-1}
\end{document}